%% file: main.tex
\documentclass[acmsmall,screen]{acmart}

\usepackage{xspace}
\usepackage{enumitem}
\usepackage{booktabs, multirow} %
\usepackage{soul}%
\usepackage[ruled, vlined, linesnumbered, commentsnumbered, longend]{algorithm2e}
\usepackage{graphicx}
\usepackage{subfigure}
\usepackage{wrapfig}
\usepackage[normalem]{ulem}
\definecolor{applegreen}{rgb}{0.55, 0.71, 0.0}
\usepackage{tcolorbox}
\usepackage{hyperref}
\hypersetup{linkcolor=black,citecolor=black,urlcolor=RubineRed}

\input{macro}

\begin{document}

\title{\tech: White-Box Compiler Fuzzing Empowered by Large Language Models}

\author{Chenyuan Yang}
\orcid{0000-0002-7976-5086}
\affiliation{%
  \institution{University of Illinois at Urbana-Champaign}
  \city{Champaign}
  \country{USA}
}
\email{cy54@illinois.edu}

\author{Yinlin Deng}
\orcid{0000-0002-4628-4219}
\affiliation{%
  \institution{University of Illinois at Urbana-Champaign}
  \city{Champaign}
  \country{USA}
}
\email{yinlind2@illinois.edu}

\author{Runyu Lu}
\orcid{0009-0000-5261-6147}
\affiliation{%
  \institution{Huazhong University of Science and Technology}
  \city{Wuhan}
  \country{China}
}
\email{lry89757@gmail.com}

\author{Jiayi Yao}
\orcid{0000-0002-8588-4356}
\affiliation{%
  \institution{Chinese University of Hong Kong}
  \city{Shenzhen}
  \country{China}
}
\email{jiayiyao@link.cuhk.edu.cn}

\author{Jiawei Liu}
\orcid{0000-0001-7122-8625}
\affiliation{%
  \institution{University of Illinois at Urbana-Champaign}
  \city{Champaign}
  \country{USA}
}
\email{jiawei6@illinois.edu}

\author{Reyhaneh Jabbarvand}
\orcid{0000-0002-0668-8526}
\affiliation{%
  \institution{University of Illinois at Urbana-Champaign}
  \city{Champaign}
  \country{USA}
}
\email{reyhaneh@illinois.edu}

\author{Lingming Zhang}
\orcid{0000-0001-5175-2702}
\affiliation{%
  \institution{University of Illinois at Urbana-Champaign}
  \city{Champaign}
  \country{USA}
}
\email{lingming@illinois.edu}

\setcopyright{rightsretained}
\acmDOI{10.1145/3689736}
\acmYear{2024}
\acmJournal{PACMPL}
\acmVolume{8}
\acmNumber{OOPSLA2}
\acmArticle{296}
\acmMonth{10}
\acmSubmissionID{oopslab24main-p305-p}
\received{2024-04-06}
\received[accepted]{2024-08-18}

\begin{abstract}

Compiler correctness is crucial, as miscompilation can falsify program behaviors, leading to serious consequences over the software supply chain.
In the literature, fuzzing has been extensively studied to uncover compiler defects.
However, compiler fuzzing remains challenging:
Existing solutions focus on black- and grey-box fuzzing, which generates test programs without sufficient understanding of internal compiler behaviors.
As such, they often fail to construct test programs to exercise intricate optimizations.
Meanwhile, traditional white-box techniques, such as symbolic execution, are computationally inapplicable to the giant codebase of compiler systems.
Recent advances demonstrate that Large Language Models (\llm{s}) excel in code generation/understanding tasks and even have achieved state-of-the-art performance in black-box fuzzing.
Nonetheless, guiding \llm{s} with compiler source-code information remains a missing piece of research in compiler testing.

To this end, we propose \tech, the first white-box compiler fuzzer using \llm{s} with source-code information to test compiler optimization{, with a spotlight on detecting deep logic bugs in the emerging deep learning (DL) compilers}. 
\tech adopts a {multi-agent} framework:
\emph{(i)} {an \llm{-based} analysis agent} examines the low-level optimization source code and produces requirements on the high-level test programs that can trigger the optimization;
\emph{(ii)} {an \llm{-based} generation agent} produces test programs based on the summarized requirements.
Additionally, optimization-triggering tests are also used as feedback to further enhance the test generation prompt on the fly.
Our evaluation on {the three most popular DL compilers (\ie{} \ptinductor, \tfxla, and \tflite)} shows that \tech can generate high-quality test programs to exercise deep optimizations requiring intricate conditions, practicing up to 8 times more optimizations than state-of-the-art fuzzers. 
To date, \tech has found in total \totalbug bugs for the compilers under test, with \newbug confirmed as previously unknown and \fixbug already fixed.
{Notably, \tech has been recently acknowledged by the \pt team, and is in the process of being incorporated into its development workflow.}
{Finally, beyond DL compilers, \tech can also be adapted for compilers in different domains, such as \llvm, where \tech has already found multiple bugs.}

\end{abstract}




\maketitle

\section{Introduction}
\input{intro}

\section{Background and Related Work}

\input{background}

\section{Design}

\input{approach}

\section{Implementation}\label{sec:impl}
\input{implement}

\section{Evaluation}
\input{exp-setup}

\section{Result Analysis}
\input{result}

\section{Discussion}

\input{discussion}

\section{Conclusion}

We present \tech, the first practical white-box compiler fuzzer to test compiler optimizations.
\tech adopts a {multi-agent} design: 
an analysis \llm{} reads through the implementation code of compiler optimizations and summarizes desired patterns of test programs, 
with which a generation \llm{} is then prompted to efficiently and continuously synthesize meaningful test programs to exercise corresponding optimizations.
Our evaluation shows that \tech is effective in testing {the emerging \dl compilers and is also adaptable to the conventional C/C++ compilers}.
To date, \tech has found in total \totalbug bugs {for \dl compilers}, with \newbug confirmed as previously unknown and \fixbug already fixed.

\section*{Data-Availability Statement}
The artifact of \tech is available at \url{https://github.com/ise-uiuc/WhiteFox}.

\begin{acks}
We thank Chunqiu Steven Xia for providing the help and resources to run some experiments.
This work was partially supported by NSF grant CCF-2131943 and Kwai Inc.
This project is supported, in part, by funding from \href{http://www.twosigma.com/}{Two Sigma Investments, LP}. Any opinions, findings, and
conclusions or recommendations expressed in this material are those of the authors and do not necessarily reflect
the views of Two Sigma Investments, LP. 

\end{acks}

\bibliographystyle{ACM-Reference-Format}
{\footnotesize 
\bibliography{main}}

\end{document}

%% file: macro.tex
\newcommand{\revision}[1]{{#1}}

\newcommand{\eg}{\emph{e.g.,}\xspace}
\newcommand{\ie}{\emph{i.e.,}\xspace}

\newcommand{\CodeIn}[1]{{\small \texttt{#1}}}
\newcommand{\Comment}[1]{}

\newcommand{\tech}{\textsc{WhiteFox}\xspace} %

\newcommand{\llm}{{LLM}\xspace} 
\newcommand{\gpt}{{GPT}\xspace} 
 
\newcommand{\starcoder}{{StarCoder}\xspace} 

\newcommand{\dl}{{DL}\xspace} 
\newcommand{\NL}{{NL}\xspace} 
 
\newcommand{\codeformat}{{pseudo-code}\xspace} 
 
\newcommand{\pt}{{PyTorch}\xspace} 
\newcommand{\tf}{{TensorFlow}\xspace} 
\newcommand{\inductor}{{Inductor}\xspace} 
\newcommand{\ptinductor}{\pt{} Inductor\xspace} 
\newcommand{\tfxla}{\tf-XLA\xspace}
\newcommand{\tflite}{\tf{} Lite\xspace}
\newcommand{\llvm}{LLVM\xspace}

\newcommand{\cpp}{{C++}\xspace} 
\newcommand{\python}{{Python}\xspace} 

\newcommand{\parabf}[1]{\noindent\textbf{#1}}

\newcommand{\batchvalue}{10\xspace}

\newcommand{\titanfuzz}{\textsc{TitanFuzz}\xspace}
\newcommand{\fuzzall}{\textsc{Fuzz4All}\xspace}
\newcommand{\nnsmith}{\textsc{NNSmith}\xspace}

\newcommand{\yarpgen}{YARPGen\xspace} %
\newcommand{\csmith}{Csmith\xspace}
\newcommand{\grayc}{\textsc{GrayC}\xspace}
\newcommand{\polyglot}{\textsc{PolyGlot}\xspace}
\newcommand{\tzer}{\textsc{Tzer}\xspace}
\newcommand{\qsym}{\textsc{qsym}\xspace}
\newcommand{\tvm}{TVM\xspace}

\newcommand{\testpilot}{TestPilot\xspace}
\newcommand{\codamosa}{\textsc{CodaMosa}\xspace}
\newcommand{\teco}{\textsc{TeCo}\xspace}

\newcommand{\totalbug}{101\xspace}

\newcommand{\fixbug}{70\xspace}
\newcommand{\highpriobug}{10\xspace}
\newcommand{\newbug}{92\xspace}

\newcommand{\pttotalbug}{79\xspace}
\newcommand{\pttotalpercent}{78.2\xspace}
\newcommand{\ptfixbug}{68\xspace}
\newcommand{\ptfixpercent}{86.1\xspace}

\newcommand{\rqmix}{WF-Mix\xspace}
\newcommand{\rqcode}{WF-Code\xspace}
\newcommand{\rqnl}{WF-NL\xspace}
\newcommand{\rqimpl}{WF-Impl\xspace}
\newcommand{\rqstnl}{WF-SC\xspace}

\newcommand{\rqrand}{\revision{WF-Naive}\xspace}
\newcommand{\rqnofb}{WF-No-Feedback\xspace}

\newcommand{\ReqGenerate}{Requirement Summarization\xspace}
\newcommand{\reqgenerate}{requirement summarization\xspace}

%% file: intro.tex
\begin{figure*}[!t]
    \centering
\includegraphics[keepaspectratio=true,width=\textwidth]{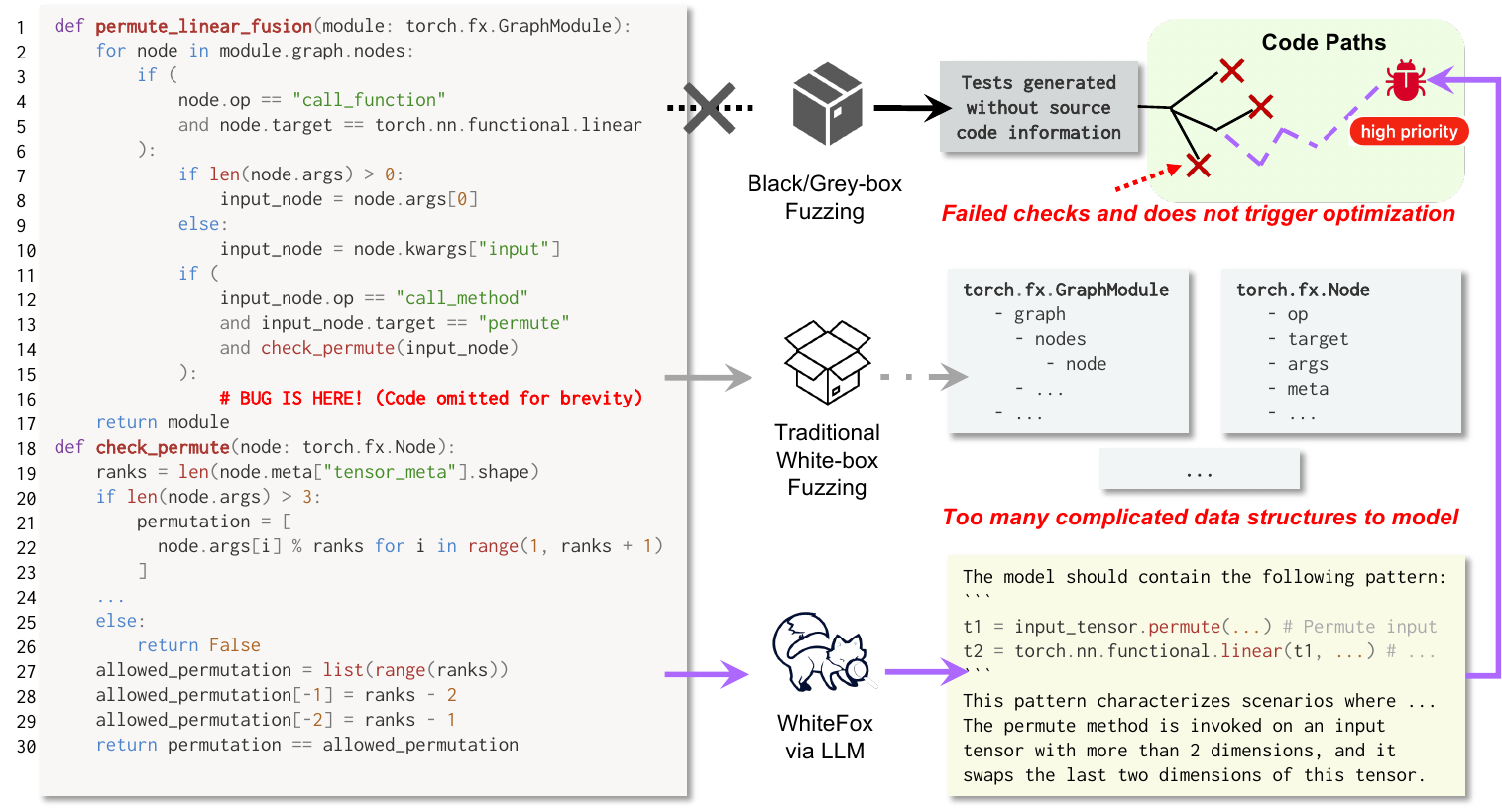}
    \caption{Motivating example.}
    \label{fig:motivation}
\end{figure*}

Modern compilers~\cite{llvm,gcc,opengl,pytorch2,Tensorflow,chen2018tvm} play a critical role in translating high-level programming languages into efficient machine code.
However, incorrect or misapplied optimizations can lead to subtle and hard-to-detect bugs, and even vulnerabilities~\cite{xu2023silent,yang2017dead,christou2023ivysyn}.
For instance, compiler misoptimizations have led to severe security vulnerabilities, \eg system hanging, memory errors, and information leaks, in the Linux kernel~\cite{xu2023silent} and safety-critical deep learning (DL) applications/systems~\cite{christou2023ivysyn,Shao2023AnES,uberkill,pei2017deepxplore}.
Given the ubiquity of compilers in software development, it is vital to ensure the correctness of compiler optimizations.
To this end, fuzzing (or fuzz testing)~\cite{sutton2007fuzzing,zeller2019fuzzing} has been applied to automatically generate a large number of test inputs, aiming to explore compiler defects~\cite{sun2016toward}.
To date, a large body of fuzzing tools has been tailored for different languages and compilers~\cite{csmith,yarpgen,nnsmith,grayc}, highlighting their success by finding a large number of real-world compiler bugs.

In the literature, researchers have proposed various fuzzing techniques to incorporate the knowledge about the system under test (SUT) during test generation~\cite{grayc,polyglot}.
They are generally classified into three categories according to the extent of SUT knowledge visible to the fuzzer: 
black-box~\cite{nnsmith,csmith,yarpgen}, 
grey-box~\cite{libfuzzer,grayc,polyglot},
and white-box fuzzing~\cite{symbolic,sen2007concolic}.
Black-box fuzzing has zero knowledge about the internal workings of the SUT and simply considers system input/interface information.
Consequently, the inputs are generated without complying with the intended structures or behaviors of the SUT.
In contrast, white-box fuzzing techniques, by inspecting the source code of the SUT, aim to synthesize test cases to exhaustively explore all possible code paths.
Grey-box fuzzing lies between black- and white-box fuzzing. 
By leveraging limited program information of the SUT (\eg code coverage), grey-box fuzzers attempt to efficiently produce tests that are more likely to exercise new program behaviors.
While in theory, we can apply all such approaches for compiler fuzzing, each approach grapples with its challenges and limitations owing to the immense complexity/scale of modern compiler systems.
For instance, the widely-used \llvm~\cite{llvm} compiler is implemented with 14M lines of code, and the popular \dl compiler, \tf~\cite{Tensorflow}, has 3.5M lines.%

\parabf{Challenges.}
Black-box fuzzing, without knowledge of the internal workings, struggles due to the intricate conditions required to trigger optimizations.
Simply generating random inputs, without any guidance, often proves impractical for reaching the deep corners of optimization logic.
For example, a recent study~\cite{polyglot} shows that the black-box fuzzer for C compilers, \csmith~\cite{csmith}, which produces test-cases through grammar-based generation, can be significantly less effective than coverage-guided fuzzing.
Grey-box fuzzing, though better informed by source code instrumentation to achieve higher code coverage than its black-box counterpart, 
frequently falls short of fully understanding the nuanced criteria required to trigger particular optimizations.
This shortcoming stems from the fact that compiler optimizations typically hinge on meeting precise and strict conditions.
Vanilla coverage-driven strategies might not navigate these specific states effectively.
Moreover, grey-box compiler fuzzing~\cite{polyglot,grayc} even 
fail to generate
semantically correct inputs, leading to the discovery of mostly front-end crash bugs. 
On the other hand, traditional white-box fuzzing, 
which relies on strict analysis of the SUT source code,
becomes daunting with modern compilers.
The sheer complexity of modern optimization techniques, combined with the vast landscape of programming paradigms and hardware targets, makes modeling all behaviors an uphill task.
For instance, symbolic execution~\cite{symbolic} executes a program by using symbolic variables in place of concrete values, enabling the systematic exploration of every potential execution path. %
However, when applied to compiler systems, it becomes infeasible to designate every variable as symbolic.
Even if such a feat were achievable, 
the million-line scale of compiler codebase inevitably leads to \textit{path explosion}, rendering the approach highly challenging.

Furthermore, traditional compiler fuzzing techniques are typically tailored to specific languages/compilers.
Yet, designing and implementing a fuzzing framework for a new compiler is both time-intensive and laborious. 
For instance, \csmith~\cite{csmith} is comprised of tens of thousands of lines of code through years of development.
Given the unique characteristics of each target language/compiler, reusing the efforts of one fuzzing implementation for a different input language/compiler often presents significant challenges.

\parabf{Motivation.}
Figure~\ref{fig:motivation} presents a motivating example of the optimization \CodeIn{permute\_linear\_fusion} in \ptinductor~\cite{pytorch2}.
This optimization fuses the \CodeIn{permute} and \CodeIn{linear} operators when the \textit{permute} method is invoked on an input tensor with more than two dimensions, specifically swapping the last two dimensions.
On the left side of the figure, we see its source code implementation.
Here, the constraints required to trigger this optimization are explicitly detailed with nested \CodeIn{if} condition statements and a helper function \CodeIn{check\_permute}.
However, when applying fuzzing techniques to test this optimization, black/grey-box fuzzing struggles to generate models that align the \CodeIn{permute} and \CodeIn{linear} operators with these constraints.
For instance, consider a scenario where a grey-box compiler fuzzer produces a model with the \CodeIn{linear} operator and covers the first \CodeIn{if-check} (Line 3-6, Figure~\ref{fig:motivation}) in this optimization.
Even if a black/grey-box fuzzer repeatedly selects this test as a seed for mutation, it is challenging to successfully mutate the model to {}invoke the \CodeIn{permute} method on a tensor—specifically, to swap its last two dimensions—where the output should then serve as the input for \CodeIn{linear}.
This is because both black-box and grey-box fuzzing are unaware that the models should include the \CodeIn{permute} and \CodeIn{linear} operators in this specific way, due to the absence of guidance from the source code implementation.
As there are thousands of operators in \pt, such fuzzers will likely choose a different operator than \CodeIn{permute} or apply \CodeIn{permute} in many other ways.
As a result, the generated models often fail validation checks and cannot activate this optimization, let alone discover deep bugs in it (Line 16, Figure~\ref{fig:motivation}).
On the other hand, though the white-box techniques have the potential to trigger this optimization theoretically, it is \textit{impractical} to apply traditional program analysis to extract constraints from the detailed source code due to the data structure complexity in compilers, \eg \CodeIn{torch.fx.GraphModule} and \CodeIn{torch.fx.Node}.
These structures are further composed of several other intricate classes with diverse attributes (\eg\space\CodeIn{args}, \CodeIn{shape}, and \CodeIn{rank}).
Additionally, the intricate constraints are often expressed in hierarchical conditions (\eg nested \CodeIn{if} statements) and even complex check functions. 
Therefore, it is extremely challenging, if not impossible, to accurately extract and express these constraints symbolically for constraint solvers, let alone apply any formal method to solve them.

\parabf{Insight.}
Can we scale white-box fuzzing to fully test optimizations for any compilers?
We address this question based on the insight that Large Language Models (\llm{s})~\cite{openai2023gpt4,chatgpt,starcoder,codex,wang2021codet5,feng2020codebert, wei2024magicoder} are pre-trained on a vast array of code spanning various programming languages. 
This broad foundation enables them to excel in comprehending and generating code across diverse languages~\cite{bubeck2023sparks}.%
Therefore, for the \CodeIn{permute\_linear\_fusion} optimization, different from typical white-box fuzzing, we can leverage \llm{s} to summarize the requirements for the models that could trigger it based on the source code information, as highlighted in the yellow text box of Figure~\ref{fig:motivation}.
Subsequently, we can leverage the generated requirement description to further prompt \llm{s} to create the corresponding inputs, which are \pt models in this case.
In our experiments, the generated tests indeed triggered the \CodeIn{permute\_linear\_fusion} optimization and even detected a previously unknown bug in it! Notably, this bug was confirmed by developers and labeled as \textit{high-priority}.

\parabf{Proposal.} We present \tech, the first \textit{white-box} compiler fuzzing approach via Large Language Models (\llm{s}) to fully test the core optimization modules in {DL compilers, which represent the fastest-evolving segment in the field of compilers}.
As discussed, existing approaches to white-box testing cannot scale to model the behavioral information of complex compiler systems.
Therefore, the key idea of \tech is to leverage \llm{s} to automatically infer the requirements of test programs that can trigger the compiler optimizations based on their source code implementation.
\llm{s}, having been pre-trained on an extensive corpus comprising natural languages and a variety of programming languages, possess the ability to comprehend and succinctly summarize optimization source code.
The input to \tech is the source code that implements compiler optimizations.
First, {an \llm{-based} analysis agent} automatically {analyzes and summarizes} the testing requirements for triggering optimizations.
Subsequently, {an \llm{-based} generation agent produces} numerous meaningful test programs guided by the generated requirements.
To generate test programs that can directly exercise corresponding optimizations, 
\tech further employs a feedback-loop mechanism that uses optimization-triggering tests as few-shot examples to guide future test generation.%

\parabf{Summary.} This work makes the following contributions:
\begin{itemize}[noitemsep, leftmargin=*, topsep=2pt]
    \item \parabf{Novelty.} We introduce a new dimension of white-box compiler fuzzing by using \llm{s} as both optimization source code analyzers and test input generators.
    To our best knowledge, this work is the first to demonstrate that \llm{s} can transform the low-level implementation information into the corresponding high-level test programs, making it practical to employ the white-box fuzzing techniques to test {complex DL compilers}. 
    {Furthermore, beyond DL compiler testing, \tech can also be adapted for white-box fuzzing of other compilers, and even complex, real-world software systems in general, inspiring future work in this promising direction.
    }
    \item \parabf{Approach.} While our approach is general and applicable to various compiler systems, we implement \tech as a practical fuzzer for {the three most popular DL compilers, \ptinductor, \tfxla, and \tflite.}
    We utilize \gpt{4}~\cite{openai2023gpt4} as an analysis agent to summarize the requirements based on the source code, and \starcoder~\cite{starcoder} as a generation agent to create diverse test inputs.
    Our artifact is available at \url{https://github.com/ise-uiuc/WhiteFox}.
    \item \parabf{Study.} We extensively compared \tech with state-of-the-art fuzzers on the target \dl compilers. Our result shows that \tech can practice 2.5x more optimizations than the baselines. By now, \tech detects \emph{\totalbug} bugs, with \emph{\newbug} already confirmed as previously unknown and \emph{\fixbug already fixed}.
    Of these, 10 bugs are labeled as \emph{high-priority} by the developers.
   {
    \item \parabf{Impact.} \tech has been acknowledged by the official \pt team and is currently being integrated into their development pipeline, demonstrating the real-world applicability of our approach. Furthermore, our generality study adapts \tech for testing \llvm and reveals multiple bugs in this popular C/C++ compiler, showing the broader impact of our work beyond DL compilers.}
\end{itemize}

%% file: background.tex
Fuzzing has been extensively studied for testing both traditional compilers and the emerging deep learning (\dl) compilers.
In this section, we first review related work for compiler fuzzing without \llm{s}.
We then discuss the recent advances of \llm-based testing.

\subsection{Compiler Fuzzing}

{
The main challenge to compiler fuzzing is to synthesize and diversify syntactically and semantically valid programs as the input to the compiler.
In the literature of traditional compiler fuzzing,
\emph{grammar-based} techniques aim to synthesize syntactically valid random programs via generation rules that comply with the language grammar.
Such techniques have been widely used for fuzzing compilers of programming languages including C/C++ (\eg \csmith~\cite{csmith} and \yarpgen~\cite{yarpgen}), JavaScript (\eg LangFuzz~\cite{holler2012fuzzing} and jsfunfuzz~\cite{jsfunfuzz}), and Python (\eg PyRTFuzz~\cite{li2023pyrtfuzz}).
However, grammar-based approaches often require massive engineering efforts to implement rules that ensure the validity of the generated programs~\cite{csmith} and may still fail to synthesize realistic yet complicated test programs.
Therefore, \emph{mutation-based} techniques~\cite{le2014compiler, donaldson2017automated,le2015finding, zhang2017skeletal} propose to mutate valid seed programs for generating new input programs
that can exercise deeper code paths in the compiler.
Besides traditional compiler fuzzing, various fuzzing techniques have also been proposed for the emerging \dl compilers.
\dl compilers compile \dl models, whose generation requires valid compositions of tensor operations.
In early work, \dl models are either directly curated from existing open-source models~\cite{cradle} or created using simple shape-preserving operators~\cite{lemon,audee}.
To diversify operators in model generation,
recent work explicitly define constraints for operator constructions either manually~\cite{nnsmith,muffin} or automatically~\cite{neuri}. 
}

{
Though comprehensive, most of the work discussed above is still black-box fuzzers.
Consequently, their generated test programs often fail to practice the internal nuances of intricate compiler behaviors, especially in the important optimization passes.
Therefore, recent grey-box compiler fuzzers~\cite{polyglot,tzer,grayc} have been proposed to integrate code coverage guidance to discover interesting input programs that can explore deeper compiler behaviors.
\polyglot~\cite{polyglot} proposes to integrate coverage feedback into its fuzzing loop to test compilers at the intermediate representation (IR) level.%
\tzer~\cite{tzer} proposes a coverage-directed fuzzing approach that jointly mutates both IR files and optimization passes, specifically targeting the \tvm{} \dl compiler~\cite{chen2018tvm}.
More recently, \grayc~\cite{grayc} combines coverage feedback with mutation operators to test C compilers.

}

{
To achieve more explicit test generation,
white-box testers by theory can tailor test programs to specifically exercise certain optimization by inspecting the compiler logic from the source code.
Yet, to our best knowledge, there is hardly any competitive white-box compiler fuzzer in existence, largely due to the unmanageable scale of compiler systems which makes detailed program analysis almost impossible.
While recent hybrid fuzzing techniques~\cite{jiang2023evaluating,qsym,choi2019grey,cho2019intriguer,kim2020hfl,liu2023dsfuzz} integrate general-purpose fuzzing~\cite{mathis2020learning,gopinath2020fuzzing,holler2012fuzzing} with concolic execution~\cite{dart,cute}, showing promise in different testing domains, their effectiveness in compiler fuzzing remains limited compared to compiler-specific fuzzers~\cite{polyglot,chen2020survey,le2014compiler,tzer}.
Besides the path explosion issue due to the compiler-scale complexity, another key limitation is that hybrid fuzzing typically operates on binary inputs, overlooking the nuanced semantics present in source code, which is crucial for compiler testing.}

\subsection{\llm-based Testing}

With the recent advancements in Large Language Models (\llm{s}), there has been a surge in efforts to leverage \llm{s} for unit test generation~\cite{zhang2023survey}.
For example, \teco~\cite{teco}, built upon a fine-tuned CodeT5~\cite{wang2021codet5} model, has been introduced to aid developers in completing unit tests for the given code and context.
\revision{\testpilot\cite{testpilot} explores adaptive zero-shot test generation with Codex~\cite{codex} for testing JavaScript. MuTAP~\cite{dakhel2024effective} leverages zero-shot and few-shot learning to generate test cases using Codex and llama-2-chat, and further evaluates the generated test cases through mutation testing. ChatUniTest~\cite{chen2024chatunitest} performs conversation-driven test generation with ChatGPT~\cite{chatgpt}.}
\revision{\codamosa~\cite{codamosa} employs the capability of \llm{s} to help with the {search-based software testing (SBST)~\cite{fraser2011evosuite}} by generating unit tests for uncovered methods.}
\revision{ChatGPT-SBST~\cite{tang2024chatgpt} performs a comprehensive study on using ChatGPT~\cite{chatgpt} for SBST.}

While \llm-based unit test generation techniques incorporate source code, they are specifically designed for particular modules (\eg functions/classes) of the projects under test.
In contrast, our approach targets the important problem of compiler testing at the entire system level (a totally different problem from unit testing), and relies solely on the source code implementation of optimizations as \textit{guidance} for fuzzing.
Also, unit test generation demands detailed and comprehensive information about the module being tested, including all associated data structures, while our guidance can be imprecise, partial, or even incomplete.
In addition, we are the first to leverage \llm{s} to bridge the gap between the low-level optimization implementation and the high-level input programs for compilers.
Lastly, \llm-based unit test generation can only assist developers and requires further manual refinement, because generating reliable unit-test oracles remains notably difficult or infeasible for even the current best \llm{s}~\cite{palm,openai2023gpt4}.
In contrast, \tech directly leverages various automated oracles for traditional compiler fuzzing, \eg differential testing~\cite{mckeeman1998differential}.

Beyond unit test generation, applying \llm{s} to fuzz software systems is another emerging trend~\cite{xia2023universal, deng2023large, titanfuzz, sunsmt, huang2024large, yang2023kernelgpt,meng2024large,ou2024mutators}.
For example, \titanfuzz~\cite{titanfuzz} demonstrates for the first time that modern \llm{s} can be directly leveraged to perform both grammar- and mutation-based fuzzing of real-world systems.
\revision{Recently, \fuzzall~\cite{xia2023universal} has also demonstrated that LLMs can serve as the universal fuzzers for various types of software systems.}
\revision{However, these approaches often neglect white-box information, specifically the source code implementation of the software under test. Consequently, they tend to concentrate on the compiler's front-end rather than the intricate logic within the middle and back-end components.}
In contrast, \tech is the first white-box compiler fuzzing approach, which leverages low-level optimization implementations for guided fuzzing, and has been demonstrated to outperform the 
 recent \titanfuzz (\S~\ref{subsec:baseline}).%

%% file: approach.tex
\begin{figure*}[t]
    \centering
    \includegraphics[keepaspectratio=true,width=0.95\textwidth]{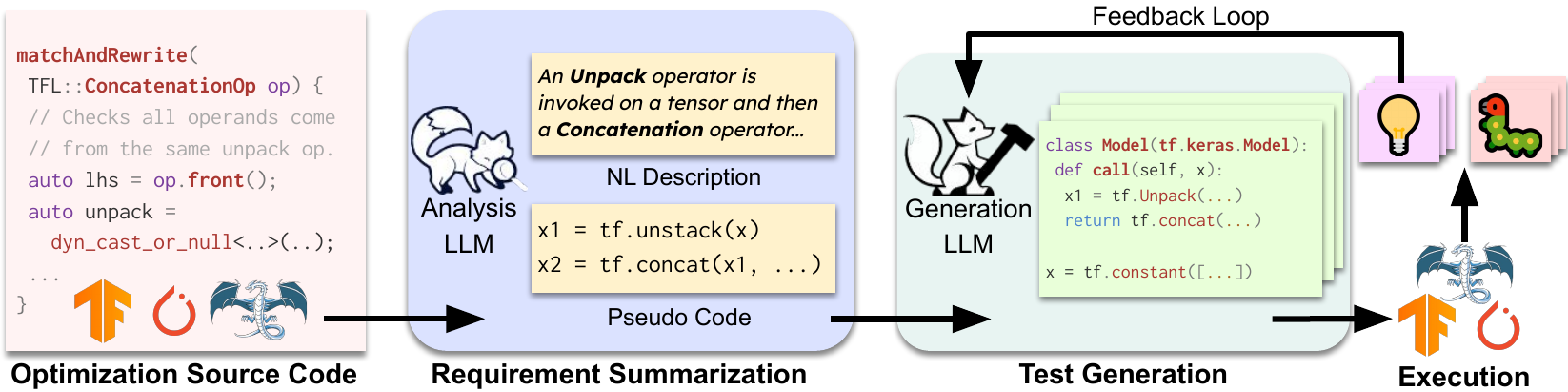}
    \caption{Overview of \tech.}
    \label{fig:overview}
\end{figure*}

Figure~\ref{fig:overview} depicts the overview of \tech, consisting of three main components: \textit{\ReqGenerate}, \textit{Test Generation}, and \textit{Feedback Loop}.
Overall, \tech takes the \textit{source code} of optimization passes from the tested compiler as inputs and generates high-quality test programs via \llm{s}. 
To that end, during the \textit{\ReqGenerate} phase, an \textit{analysis \llm} is used to summarize the requirements of the test programs to trigger the optimization by examining its source-code implementation (\S~\ref{subsec:nl}).
Next, the analyzed requirements are used to prompt a \textit{generation \llm}, which synthesizes test programs to specifically practice corresponding optimizations (\S~\ref{subsec:test}).
The generated test programs are then compiled and executed, and \tech observes whether they can indeed activate the corresponding optimizations via instrumentation.
If a test program triggers any optimization, it will be incorporated into \tech{'s} feedback loop as an example to guide the \textit{generation \llm} towards more optimization-targeted generation in subsequent iterations (\S~\ref{subsec:feedback}). 
To detect compiler bugs, every test program is executed with test oracles including result inconsistency, as well as compile- and execution-time crash (\S~\ref{subsec:oracle}).

Notably, \tech employs a {multi-agent} framework: 
\emph{(i)} an \llm{-based} analysis agent is prompted to infer the conditions for triggering optimizations by inspecting the implementation code; 
and
\emph{(ii)}
an \llm{-based} generation agent is prompted to create a large number of meaningful test programs.
This design allows us to balance the trade-off between the costs and benefits that different \llm{s} provide.
For example, we can let the analysis \llm be one with broad knowledge and reasoning ability (in both natural language and code)
and let the generation \llm be one that is specialized for efficient program generation.

\subsection{\ReqGenerate}
\label{subsec:nl}

While it is possible to directly generate tests from the optimization source code with \llm{s}, it does not perform well (as shown in our evaluation \S~\ref{subsec:rqstudy}).
This is because the optimization source code has the following two undesirable features: \emph{(i)} The implementation source code for the optimization is often lengthy. It often contains \textbf{redundant} information like detailed comments explaining the code behavior. Furthermore, it may contain code that is \textbf{unrelated} to the characteristics of triggering input, such as implementation details (e.g., using data structures like \CodeIn{std::<vector>} to gather operands), the IR transformation after the optimization is triggered, and error logging. Such long source code not only makes it time-consuming for \llm{s} to process but also adds to the difficulty for \llm{s} to understand, as \llm{s} are known to struggle with long context ~\cite{liu2023lost}.
Additionally, some source code may exceed the limited context window available to \llm{s}. \emph{(ii)} The implementation code is written at a \textbf{low level}, involving large quantities of domain-specific modules, IRs, and helper functions. For example, from Figure~\ref{fig:overview}, we can observe that the optimization source code relies on \tflite-specific module \CodeIn{TFL}, which contains various low-level operators (\eg{} \CodeIn{TFL::ConcatenationOp} and \CodeIn{TFL::UnpackOp}) and functions (\eg{} \CodeIn{dyn\_cast\_or\_null}).
These IRs are translated from the input programs written at a higher level. They are meant for internal use within a compiler, and none of them are visible at the high level.%

To address the above challenges, we propose to leverage an analysis \llm{} to provide \textbf{a clear, concise, and comprehensible summary} of how the optimization can be triggered, which will assist in subsequent test generation.
We considered two types of summary formats: \textbf{natural language} and \textbf{pseudo-code}.
Natural language description can be more concise than the original optimization source code, as it only summarizes the key aspects of requirements and ignores redundant or unrelated information. Furthermore, it is much more comprehensible to \llm{s}, since \llm{s} have been trained on enormous natural language data. On the other hand, pseudo-code can concisely describe certain patterns (especially when there are long sequences of function calls) and is closer to the structure of tests that we want to generate ultimately.
Additionally, we map the low-level implementation to a high-level summarization using natural language or pseudo code resembling user code.
For instance, \CodeIn{TFL::ConcatenationOp} is a low-level \tflite IR used in the implementation source code of \tflite optimizations (in Figure~\ref{fig:overview}), which corresponds to the high-level \tf public API \CodeIn{tf.concat}, and semantically means joining data from multiple input tensors.
In our natural language description/pseudo code, instead of using the low-level \CodeIn{TFL::ConcatenationOp}, we directly use ``concatenation operator''/\CodeIn{tf.concat} to summarize.

While both natural language and pseudo-code can help shorten the context and remove low-level information to avoid confusing the \llm, each has its own unique strengths.
For example, the \CodeIn{permute\_linear\_fusion} optimization from \ptinductor requires that \textit{``the tensor method permute is invoked first, and then the torch.nn.functional.linear function is invoked on the permuted tensor''}.
Such a \NL description is not as straightforward as the \codeformat format for this case.
Specifically, \textit{``the tensor method permute''} could be simply represented as \CodeIn{input\_tensor.permute(...)} in \codeformat.
However, for the description \textit{``it swaps the last two dimensions of this tensor''}, it is pretty challenging to leverage \codeformat to describe it clearly and briefly.
To combine and maximize their strengths, we adopt a hybrid format that blends \NL and \codeformat to describe the requirements for triggering optimization, rather than relying solely on either format.
This mixed format provides the analysis \llm{} with greater flexibility to utilize \NL and/or \codeformat as needed for each component of requirements, resulting in a more expressive and higher-quality summarization.
Our experimental findings (\S~\ref{subsec:rqstudy}) also support that the mix of \NL and \codeformat achieves the best performance in this task.

Despite all these aforementioned benefits of a high-quality optimization summary, it is worth noting that the \reqgenerate is a very comprehensive and challenging task, even for domain experts.
First, it requires understanding the logic of the implementation code and rephrasing it with semantic-preserving natural language or pseudo-code.
Second, the mapping from low-level implementation code to high-level information necessitates a broad background knowledge of the corresponding programming language and compiler.
This ranges from understanding general technical terminology (\eg ``variable arguments'' and ``tail calls''), to in-depth domain-specific knowledge (\eg from low-level LLVM IRs to high-level C++ grammar).
As demonstrated in our evaluation (\S~\ref{subsec:rqstudy}), the most powerful \llm to date (namely \gpt{4}) is capable of performing this challenging analysis process (while the current open-source models, such as \starcoder, still lag far behind). This capability stems from its extensive training on vast datasets, enabling it to gain a broad knowledge and implicit understanding of various programming languages and systems. Additionally, its proficiency in performing these domain-specific analyses in our work likely results from its training on the source code of these open-source compiler systems. %
\begin{figure}[t]
    \centering
  \subfigure[Prompt template for \reqgenerate.]{
    \includegraphics[width=0.47\textwidth]{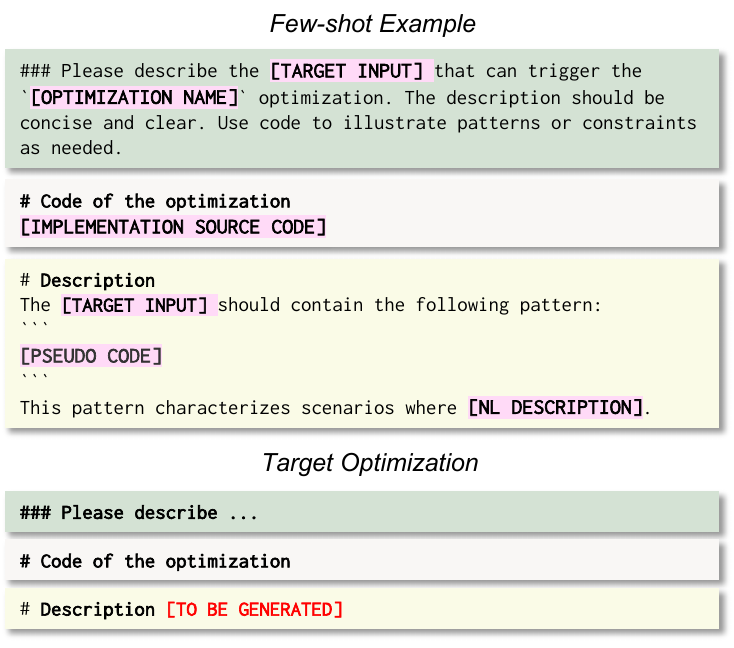}
    \label{fig:nl-general}
  }
  \subfigure[Requirement summarization prompt in \ptinductor.]{
    \includegraphics[width=0.47\textwidth]{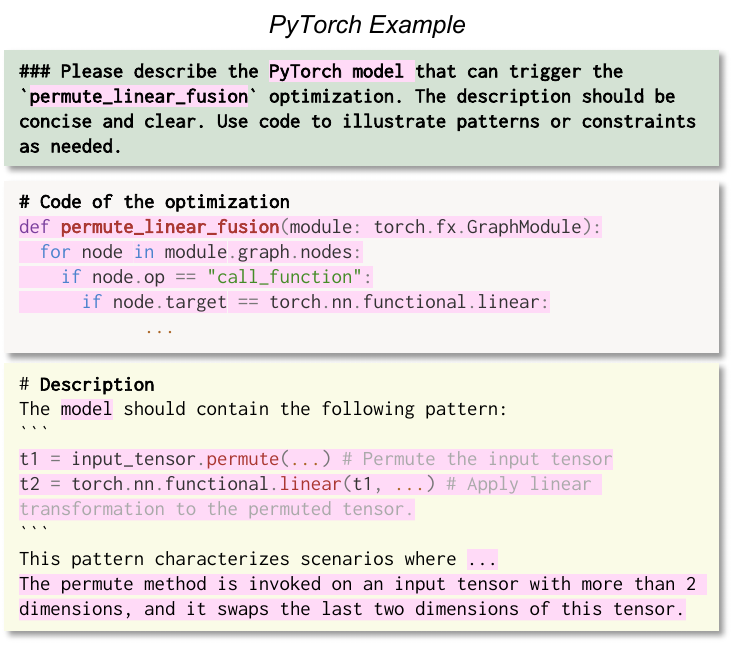}
    \label{fig:nl-generation}
  }
    \caption{Requirement summarization prompt in \tech.}
    \label{fig:nlgen-prompts}
\end{figure}

Therefore, \tech first leverages the analysis \llm to infer the requirement of high-level inputs that could trigger the optimization, utilizing its implementation written in the low-level source code.
More specifically, for each optimization, we use few-shot in-context learning ~\cite{fewshot} to prompt the analysis \llm to generate its trigger requirements for the inputs in the mixed format of \NL and \codeformat.
Figure~\ref{fig:nl-general} presents the \textit{general} few-shot prompt template used to summarize requirements for optimizations in target compilers.
This prompt template starts with the instruction \textit{``Please describe the \CodeIn{[TARGET INPUT]} that can trigger the \CodeIn{[OPTIMIZATION NAME]} optimization...''}, where \CodeIn{[TARGET INPUT]} is the input format specific to the target compiler.
Then it is followed by the source code of the optimization implementation and concludes with the description of requirements that the input should fulfill.
Note that the description is in the mixed format of \NL and \codeformat, consisting of \CodeIn{[PSEUDO CODE]} and \CodeIn{[NL DESCRIPTION]}.
The target optimization has the same structure as the few-shot examples, but its requirements field is left empty, awaiting generation by the \llm.
Take the prompt of \reqgenerate in \ptinductor as an example, whose few-shot prompt is shown in Figure~\ref{fig:nl-generation}.
The expected input format for \ptinductor is a \pt model; therefore, \CodeIn{[TARGET INPUT]} is populated with ``\pt model''.
Following this, the source code for the example optimization, \CodeIn{permute\_linear\_fusion}, is provided.
Finally, example descriptions are given in the mix of \codeformat and \NL formats to outline the constraints necessary to trigger the example optimization.
Note that such few-shot examples can guide the analysis \llm to generate desired output formats.
For instance, the example description (emphasized with a yellow box in Figure~\ref{fig:nl-generation}) provides the \llm with a clearer illustration of expected formats.
Furthermore, they facilitate the learning process of analysis \llm by providing the example mappings from low-level optimization implementation to high-level input program requirements.

\newcommand{\llmprob}{\mathcal{P}}
\newcommand{\optname}{O}
\newcommand{\optinstruct}{I}
\newcommand{\optsrc}{C}
\newcommand{\optdesc}{R}
\newcommand{\kshotnlgen}{K}
\newcommand{\examples}{E_{\kshotnlgen}}
\newcommand{\targetopt}{t}
More formally, let $\llmprob_{A}$ be the analysis \llm that models the probability of token sequences. It takes the following types of information as inputs: \emph{(i)} $\optinstruct_i$, the instruction on summarizing an optimization $\optname_i$; \emph{(ii)} $\optsrc_i$, the source code implementation of $\optname_i$; \emph{(iii)} $\optdesc_i$, the summarized trigger pattern or requirement of the optimization $\optname_i$. Let $\examples$ be the few-shot prompt prefix consisting of $\kshotnlgen$ example optimizations: $\examples=(\optinstruct_{1}, \optsrc_{1}, \optdesc_{1})\circ(\optinstruct_{2}, \optsrc_{2}, \optdesc_{2})\circ \ldots \circ (\optinstruct_{\kshotnlgen}, \optsrc_{\kshotnlgen}, \optdesc_{\kshotnlgen})$. Let $\optname_{\targetopt}$ be the target optimization. The probability distribution of the generated requirement $\optdesc_{\targetopt}$ can be defined as $\llmprob_{A}(\optdesc_{\targetopt} \mid \examples, \optinstruct_{\targetopt}, \optsrc_{\targetopt})$.

\subsection{Test Generation}
\label{subsec:test}

\begin{figure}[t]
    \centering
  \subfigure[Prompt template for test generation]{
    \includegraphics[width=0.47\textwidth]{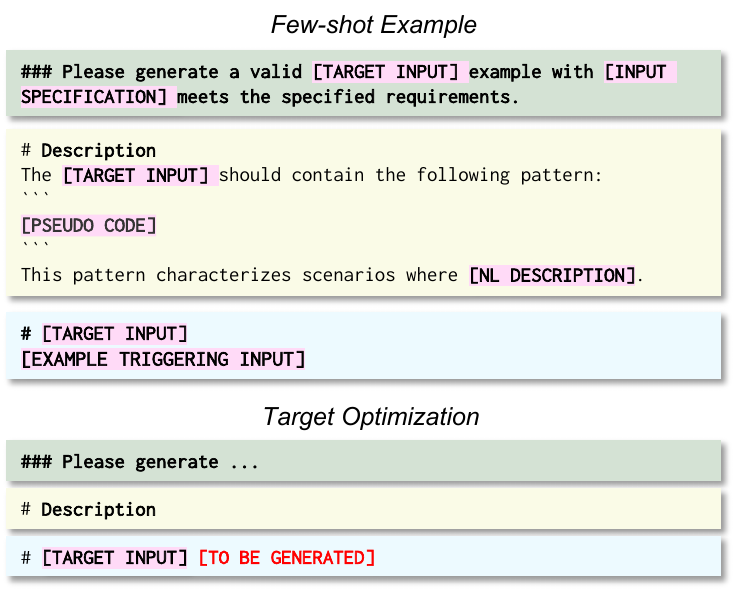}
    \label{fig:test-general}
  }
  \subfigure[Test generation prompt in \ptinductor.]{
    \includegraphics[width=0.47\textwidth]{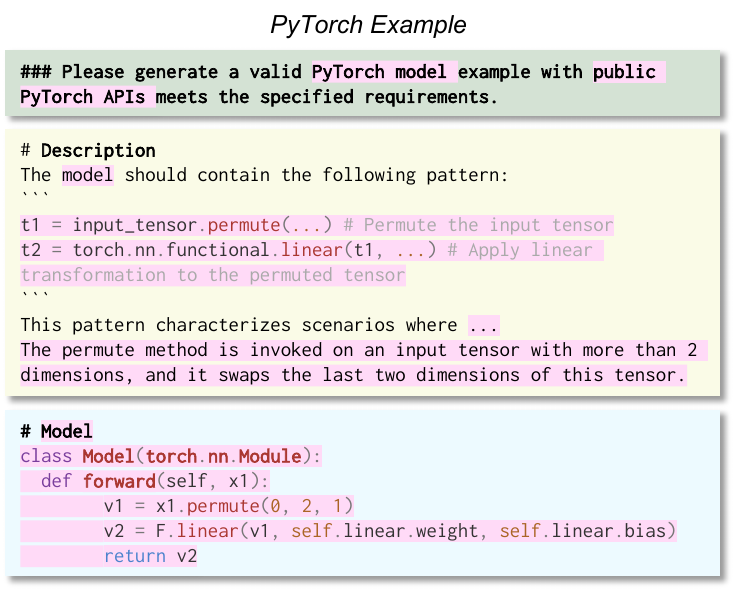}
    \label{fig:test-generation}
  }
    \caption{Test generation prompt in \tech.}
    \label{fig:testgen-prompts}
\end{figure}

By utilizing the requirement description of the optimization, \tech employs the power of \llm to generate test inputs that can effectively trigger the corresponding optimization for bug detection. 
Similar to \reqgenerate, we leverage few-shot in-context learning to generate the test inputs specific to each optimization based on the requirements. %
Figure~\ref{fig:test-general} shows the \textit{general} prompt template used in \tech for test generation.
In the template, the structure of few-shot examples comprise an instruction, specifically: \textit{``Please generate a valid \CodeIn{[TARGET INPUT]} example with \CodeIn{[INPUT
SPECIFICATION]} meets the specified requirements.''}
Then, it details the requirements for optimization activation and concludes with an illustrative input that practices the \emph{example} optimization.
Following the few-shot examples, the target optimization has a similar structure while the test input to it is \emph{to be generated} by the \llm.
Figure~\ref{fig:test-generation} shows the test generation prompt used in \ptinductor.
The format of test input to the \ptinductor is a \pt model (\CodeIn{[TARGET INPUT]}) with public \pt APIs (\CodeIn{[INPUT SPECIFICATION]}).
Next, we specify the requirements for the test inputs to activate the optimization \CodeIn{permute\_linear\_fusion}.
This is complemented by an illustrative model that can trigger this optimization.%
The provided few-shot examples aid the \llm in generating the test in the desired format, such as a \CodeIn{torch.nn.Module} composing public \pt APIs.
Furthermore, the example could help the \llm learn the relationship between the requirement description of the optimization and the corresponding test input that can trigger it.
In Figure~\ref{fig:test-generation}, the model input highlighted in a blue box (\textit{\# Model}) contains the code \CodeIn{x1.permute(0, 2, 1)}.
This corresponds to the requirement: \textit{"The permute method is invoked on an input tensor with more than 2 dimensions, and it swaps the last two dimensions of this tensor"}.
Such examples elucidate for the \llm: \emph{(i)} the meaning of "permute method is invoked on an input tensor"—which should not be \CodeIn{torch.permute(x1, ...)}, and \emph{(ii)} the implication of "swaps the last two dimensions of this tensor"-which is \CodeIn{permute(0, 2, 1)} for the tensor with three dimensions.

\newcommand{\opttest}{T}
Again, let $\llmprob_{G}$ be the generation \llm that models the probability of token sequences. Recall that $I$ represents the summarization instruction and $R$ denotes the requirement for triggering an optimization. Let $\examples$ be the few-shot prompt prefix consisting of $\kshotnlgen$ example optimizations: $\examples=(\optinstruct_{1}, \optdesc_{1}, \opttest_{1})\circ(\optinstruct_{2}, \optdesc_{2}, \opttest_{2})\circ \ldots \circ (\optinstruct_{\kshotnlgen}, \optdesc_{\kshotnlgen}, \opttest_{\kshotnlgen})$, where $\opttest_{i}$ is a valid test input that triggers the optimization $\optname_i$. Let $\optname_{\targetopt}$ be the target optimization. The probability distribution of the generated test $\opttest_{\targetopt}$ can be defined as $\llmprob_{G}(\opttest_{\targetopt} \mid \examples, \optinstruct_{\targetopt}, \optdesc_{\targetopt})$.

\subsection{Feedback Loop}
\label{subsec:feedback}

The aim of utilizing the optimization implementation to guide the \llm in test generation (\S~\ref{subsec:test}) is to produce diverse tests that can effectively exercise the target optimization, with the ultimate aim of uncovering potential bugs within it.
While the prompt in Figure~\ref{fig:testgen-prompts} contains example optimizations from the same compiler under test (and their trigger tests), these few-shot examples lack specific guidance for the target optimization.

The motivation behind incorporating a feedback loop is that the generated tests themselves are valuable feedback from the SUT, and can serve as guidance for fuzzing a target optimization in future iterations.
In each iteration, when newly generated tests are discovered for the target optimization, we collect them as candidates for few-shot examples for future test generation prompts.
By incorporating such successful triggering tests into the prompt, we enhance a targeted guidance to the generation \llm, enabling it to produce more inputs that trigger the target optimization.

\begin{algorithm}[t]
\small
\caption{\revision{Example Selection Algorithm Inspired by Thompson Sampling}}
\label{algo:tpsample}
\DontPrintSemicolon
\SetKwProg{Fn}{Function}{:}{}
\SetKwData{triggertests}{TriggerTests}
\SetKwData{choosen}{ExampleTests}
\SetKwData{newtrigger}{NewTriggerTests}
\SetKwData{numtrigger}{NumTrigger}
\SetKwData{numnottrigger}{NumNotTrigger}

\SetKwFunction{Select}{\textsc{Select}}
\Fn{\Select{\triggertests, N}}{
\For{$t \in \triggertests$}{\label{algo:sample-start}
Sample $\theta_t \sim$ Beta($\alpha_t, \beta_t$)\;
}\label{algo:sample-end}
$\choosen \leftarrow argmax_t(\theta, N)$\;\label{algo:choose}
\Return{$\choosen$}\;
}

\SetKwFunction{Update}{\textsc{Update}}
\Fn{\Update{\choosen, \newtrigger, \numtrigger, \numnottrigger}}{
\For{$t_{example} \in \choosen$}{\label{algo:parent-start}
$\alpha_{example} \leftarrow \alpha_{example} + \numtrigger$\;
$\beta_{example} \leftarrow \beta_{example} + \numnottrigger$\;
}\label{algo:parent-end}
$\alpha_{avg} \leftarrow avg(\{\alpha_{example} | t \in \choosen\}])$\;\label{algo:new-start}
$\beta_{avg} \leftarrow avg(\{\beta_{example} | t \in \choosen\}])$\;
\For{$t_{new} \in \newtrigger$}{
$\alpha_{new} \leftarrow \alpha_{avg}$\;
$\beta_{new} \leftarrow \beta_{avg}$\;
}\label{algo:new-end}
$\triggertests \leftarrow \triggertests \cup \newtrigger$\label{algo:update}
}

\end{algorithm}

\begin{wrapfigure}{r}{0.49\textwidth} %
    \centering
    \includegraphics[keepaspectratio=true,width=\linewidth]{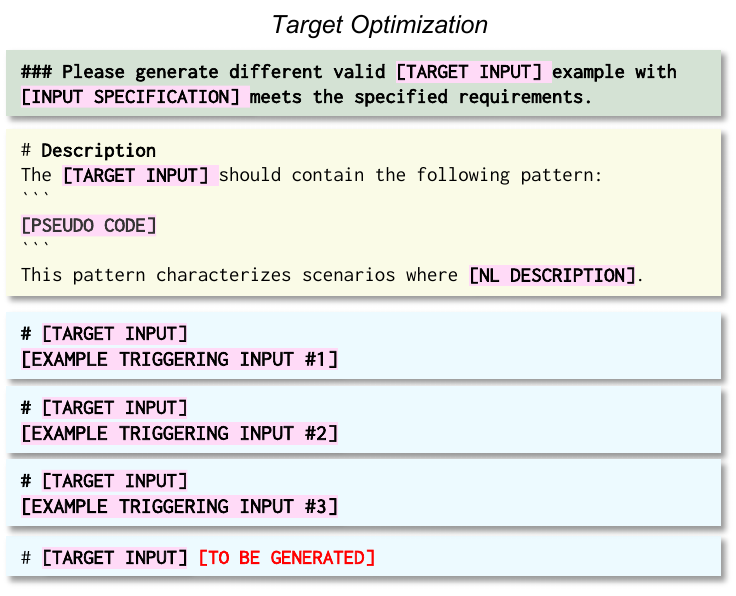}
    \caption{Prompting for test generation with feedback.}
    \label{fig:prompt-feedback}
\end{wrapfigure}

Based on the inspiration, \tech incorporates tests that have successfully triggered the corresponding optimization as supplementary examples during iterative test generation.
The target compiler is instrumented to record triggered optimizations for each test input.
If the test input successfully practices the corresponding optimization, it will be incorporated as a candidate example in the subsequent iteration of test generation for that specific optimization.
More specially, in each iteration, if there are triggering inputs for the current optimization, \tech selects several triggering inputs as examples (defaulting to 3).
These example triggering inputs will be plugged into the prompt shown in Figure~\ref{fig:prompt-feedback}, along with the instruction and requirement summary of the target optimization (same instruction and summary as the previous prompt), and will be used to generate the next batch of test inputs.
This feedback design aids the \llm in generating tests that have a higher likelihood of triggering the targeted pattern, which is demonstrated in our ablation study (\S~\ref{subsec:fbstudy}).
In the case where there are no triggering inputs for a particular optimization, \tech continues to use the initial prompt (Figure~\ref{fig:testgen-prompts}) in subsequent iterations until it finds an input capable of triggering that optimization.

To further investigate this, we observe that not all triggering examples are equally effective in guiding the \llm to generate new valuable triggering tests.
One useful signal for assessing their effectiveness is the triggering rate of the newly generated tests when we employ them as few-shot examples.
To effectively select triggering examples with an evolving knowledge of example effectiveness, it is crucial to find a balance between exploration and exploitation.
Exploration is critical because it not only allows us to evaluate under-explored options but also helps to produce a diverse set of tests for fuzzing.
On the other hand, some level of exploitation is desirable, as it enables us to fully harness the potential of effective examples.
To address this, \tech adopts an (adapted) Multi-Armed Bandit (MAB) algorithm, Thompson Sampling ~\cite{thompson1933likelihood}, as the selection strategy for triggering examples to balance the exploiting and exploration trade-off.
Each triggering example is conceptualized as an arm in the MAB framework.
The main assumption here is that each triggering example is associated with a probability representing the triggering rate, which quantifies the proportion of generated tests capable of triggering the optimization when utilizing the given example.
During the fuzzing loop, our objective is to estimate the probability associated with each triggering example, with the aim of using the most effective example to achieve more valuable triggering tests.
More specifically, following the classical Thompson Sampling algorithm ~\cite{thompson1933likelihood}, when we do not have any prior information about an arm, we choose standard beta distribution~\cite{mcdonald1995generalization} $B(1, 1)$ (or equivalently Uniform$(0,1)$) for its prior distribution. The beta distribution is parameterized by two shape parameters $\alpha > 0, \beta > 0$, which represents the number of successes and failures in historical trials.
The probability density function of beta distribution can be formally written as follows: 
$$f_{\alpha,\beta}(x)=\frac{1}{B(\alpha, \beta)}x^{\alpha-1}(1-x)^{\beta-1}$$
where $B(\alpha, \beta)$ is a constant for normalization. After drawing a new sample and observing the reward (in our case, $1$ if a generated test successfully triggers the targeted optimization and $0$ otherwise), the posterior probability can be conveniently updated by increasing $\alpha$ or $\beta$ by one, depending on whether the sample was a success or failure. More formally, if our prior belief about $X$ is represented by $f_{\alpha, \beta}$, the posterior distribution of $X$ will be updated to $f_{\alpha+1,\beta}$ or $f_{\alpha,\beta+1}$ after we observe $X=1$ or $X=0$.

Algorithm~\ref{algo:tpsample} shows the example test selection process.
Firstly, we sample $\theta_t$ from each of the posterior distributions (Line~\ref{algo:sample-start}-\ref{algo:sample-end}).
Subsequently, we opt for top-$N$ arms with the highest sampled value, which are the chosen example tests (\CodeIn{ExampleTests}) in this iteration (Line~\ref{algo:choose}).
Unlike conventional scenarios where a single arm is chosen, we simultaneously select multiple test examples at a time to construct a single few-shot prompt for generating a batch of new tests.
Consequently, when we observe the number of triggering tests among all newly generated tests, we use this information to update the posterior of each of \CodeIn{ExampleTests} (Line~\ref{algo:parent-start}-\ref{algo:parent-end}).
We next initialize the new trigger tests (\CodeIn{NewTriggerTests}) using the mean values of $\alpha$ and $\beta$ of the distribution of the \CodeIn{ExampleTests} (Line~\ref{algo:new-start}-\ref{algo:new-end}), to reduce the overhead of re-exploring the distribution of \CodeIn{NewTriggerTests} from scratch.
This comes from our assumption that the effectiveness of the new test is highly correlated with those few-shot examples, as the new test is generated by the \llm based on those specific examples, potentially inheriting valuable code patterns from such ``parent'' examples~\cite{fewshot}.
In the end, we update the pool of existing trigger tests with newly found tests (Line~\ref{algo:update}).

\subsection{Test Oracle}
\label{subsec:oracle}

Bugs are manifested in the following symptoms under \tech.

\parabf{Result inconsistency.}
During compilation, programs are iteratively transformed to semantically equivalent yet more efficient code through an array of optimization passes.
However, miscompilation can silently happen due to logical defects in the pass implementation, leading to undesired semantic inconsistency in the produced machine code.
Such semantic inconsistency can be manifested through differential testing, as is commonly used in prior compiler testing work~\cite{nnsmith,tzer,le2014compiler,le2015finding,yarpgen}.
Specifically,
for each test program that is both compilable and executable,
given the same set of  inputs to the program (if required),
we cross-check the produced outputs over the optimized and non-optimized (or minimally optimized) versions of the program.

\parabf{Crash.}
Following prior work~\cite{titanfuzz,nnsmith,csmith,grayc,le2015finding,eagle,polyglot,tzer},
it is undesirable to let the compiler and the compiled executable crash unexpectedly.
Consequently, \tech actively captures crashing signals at both the compile- and execution-time for the test program, including process aborts, segmentation faults, and unexpected internal exceptions (\eg{} \CodeIn{INTERNAL\_ASSERT\_FAILED} in PyTorch).

\revision{
To summarize, we compile each test input in two modes: with and without optimization. We consider the following conditions as bug candidates:
\begin{itemize}[noitemsep, leftmargin=*, topsep=2pt]
    \item Crashes during either optimization compilation or optimized program execution.
    \item Discrepancies in compilation status (pass/fail) between the two modes.
    \item Different program outputs between the two modes.
\end{itemize}
}

%% file: implement.tex
\parabf{Optimization collection.} 
{We start to gather optimization-specific compiler source code by specifying the relevant directories.
For example, the source code of optimization passes for \ptinductor is managed under the \CodeIn{torch/\_inductor} directory, and that for {\tfxla is mainly placed under \CodeIn{tensorflow/compiler/xla/service}}.
We next identify code fragments (\eg functions) that perform optimizations through simple keyword pattern matching.
For instance, operator fusion is an important optimization in \dl compilers and we collect the relevant functions by searching ``fusion'' or ``fuse''{}.
In addition, auxiliary functions invoked by the main optimizations are also collected since they may unveil essential conditions for activating the optimizations.
Curating and identifying optimization-relevant code fragments is required to drive \tech; however, it shall be easy for compiler developers who have a deep understanding of the code being maintained.
}

\parabf{Instrumentation.} 
{To gather the triggering information for the feedback loop (\S~\ref{subsec:feedback}), we instrument each collected optimization function by inserting a logging statement at the function entry.
As such, when compiling a test program, from the logs a sequence of activated optimization passes can be obtained. 
}

\parabf{Analysis and generation \llm{s}.}
While our approach is general and thus agnostic to the \llm{s} being employed,
our tool \tech{} is built on the state-of-the-art \gpt{4}~\cite{openai2023gpt4} and \starcoder~\cite{starcoder}.
Specifically, we utilize \gpt{4}~\cite{openai2023gpt4} as the analysis \llm{} for its recognized excellence in code comprehension and proficiency in natural language processing tasks~\cite{bubeck2023sparks}. %
           {For each optimization, we let the analysis \llm{} generate one requirement description with the temperature set to zero via the OpenAI APIs.}
           {Meanwhile, we choose \starcoder~\cite{starcoder} \revision{(\starcoder-15B)} to be the generative \llm{}, which is} an open-source model with 15.5B parameters and a context length of 8K.
In each iteration, 
           {we let \starcoder generate a batch of ten test inputs with the temperature set to one through the HuggingFace APIs~\cite{HuggingFaceWebPage}.}
The model choices allow us to balance the trade-off between the costs and benefits that different \llm{s} provide:
\emph{(i)} \gpt{4} is a powerful unified \llm (\ie with broad knowledge and reasoning ability over both natural language and code) but costly, making it suitable as the analysis \llm where its use is a one-time effort;
\emph{(ii)} \starcoder is an affordable \llm{} specialized for code and is thus suitable for efficient continuous test generation. 

\parabf{Few-shot prompting.}
For the requirement summarization and initial test generation few-shot prompt specific to each target compiler, we opt for one-shot prompting, for minimal manual efforts involved in prompt construction and affordable \llm{} context size.
To accomplish this, we select an optimization from each target compiler.
Subsequently, we manually write the requirement description and a demonstrative input test capable of triggering the optimization.
This serves as the one-shot example in the prompts for both \reqgenerate and test generation.
One exception is that \ptinductor has two distinct types of optimizations (7 utilizes a conventional optimization check function, and 54 involves a pattern matcher).%
Therefore, we separately design two prompts for each type and choose the corresponding prompt for each optimization based on its type.
For the feedback prompt, the requirement is produced by the analysis \llm{}, and the sample tests are created by the generation \llm{}. 
We use three-shot as our default setting in the feedback prompt.
\revision{Overall, constructing prompts for each target compiler is relatively straightforward. We only included a single example per compiler to illustrate the task format, requiring minimal effort. It is even easier for compiler developers who are familiar with optimization logic. Notably, such examples might already exist in test suites for many compilers. For comparison, many traditional compiler fuzzing techniques even require numerous example tests as seeds~\cite{le2014compiler,le2015finding,zhang2017skeletal}.}

%% file: exp-setup.tex
\subsection{Research Questions}

We investigate the following research questions in our experiments:

\begin{itemize}[noitemsep, leftmargin=*, topsep=0pt]
    \item {\textbf{RQ1:} How does \tech compare to state-of-the-art DL compiler fuzzers?}
    \item {\textbf{RQ2:} Are all the key components in \tech effective? }
    \item {\textbf{RQ3:} Is \tech able to detect real-world bugs?}
\end{itemize}

\subsection{Experimental Setup}
\label{sec:expsetup}
\parabf{Compilers under test.} {Our main targets are the three most popular \dl compilers: \ptinductor~\cite{pytorch2}, \tflite~\cite{tflite} and \tfxla~\cite{tfxla}, within \pt~\cite{pytorch} and \tf~\cite{Tensorflow}, two of the most widely used \dl frameworks. 
With our best effort, we collect all possible optimizations from these compilers. 
For \tfxla, whose optimization implementations tend to be lengthy, we only choose the optimizations that consist of less than 400 lines due to the limit of \llm context window size.
Table~\ref{tab:backend} lists the overview of the tested \dl compilers.
}

\begin{table}[!htp]\centering
\caption{Details of target \dl compilers.}\label{tab:backend}
\small
\begin{tabular}{l|rrrr}\toprule
            &\# Optim.  & Source lang. & Test lang. & Nightly ver. \\\midrule
\ptinductor &61 &Python & Python  & 20230509 \\
\tflite     &13 &C++    & Python  & 20230507 \\
\tfxla      &49 &C++    & Python  & 20230507 \\
\bottomrule
\end{tabular}
\end{table}

\parabf{Baselines.}
We compare \tech with state-of-the-art DL system fuzzers, including \llm-based \titanfuzz~\cite{titanfuzz} and symbolic rule-based \nnsmith~\cite{nnsmith}.
Since most optimizations are triggered by a sequence of operators, we do not include the comparison with API-level DL library fuzzers~\cite{freefuzz, docter}, which are not intended for testing optimizations.
We evaluate each baseline tool using its default configuration, \eg a 4-hour input generation time for \nnsmith.
We retain the default settings for all compared baselines because they were selected by the original authors as the optimal parameters for achieving high performance while minimizing saturation.
{

Notably, there are no practical white-box fuzzers specifically targeting \dl compilers to the best of our knowledge.
For general-purpose directed or hybrid fuzzing approaches (\eg \qsym~\cite{qsym} and \textsc{Pangolin}~\cite{huang2020pangolin}), since they are far less effective for large-scale/complicated compiler systems compared to compiler-specific techniques~\cite{polyglot,grayc,chen2020survey}, we exclude them from our baselines.
{More importantly, to the best of our knowledge, there is no directed or hybrid fuzzing approach for \dl compilers.
One possible reason could be the inherent complexity of \dl compilers, making it prohibitively difficult to craft such tools. Specifically, \dl compilers are often developed in diverse programming languages (\eg C++, Python, and CUDA) and rely heavily on various backend libraries (\eg Triton~\cite{triton}, oneDNN~\cite{onednn}, and MKL-DNN~\cite{mkldnn}).
Additionally, generating arbitrary inputs for \dl compilers is extremely difficult for general-purpose fuzzers due to dual requirements: satisfying language syntax/semantics (e.g., Python's dynamic typing and syntax checks) and tensor/operator constraints for valid computational graphs~\cite{titanfuzz,nnsmith}. 
As a result, we opt to compare with the current best techniques for fuzzing \dl compilers, \ie \titanfuzz~\cite{titanfuzz} and \nnsmith~\cite{nnsmith}.
}
}

\parabf{Ablation variants.} Multiple \tech variants are evaluated in our ablation study.
Considering that \ptinductor boasts the most optimizations, we conduct our ablation study exclusively to \ptinductor.
For requirement generation, we consider four variants: \textbf{\rqmix} (the default setting of \tech{}), \textbf{\rqnl}, \textbf{\rqcode}, and \textbf{\rqimpl}.
For each optimization, we let the generation \llm generate 100 test inputs, guided by different requirement formats.
Our default setting, \textbf{\rqmix}, describes the requirements in the mixed format of natural language and \codeformat generated by the analysis \llm.
By contrast, the requirements used in \textbf{\rqnl} (resp. \textbf{\rqcode}) are the natural language (resp. \codeformat) description extracted from the mixed format, for a fair comparison.
Besides, we also evaluate the performance of directly feeding the generation \llm with the implementation source code, \ie the \textbf{\rqimpl} variant.
Regarding the feedback loop, in addition to our default configuration, which uses feedback with Thompson Sampling, we contemplate two alternative variants: one without any feedback (\textbf{\rqnofb}) and another that incorporates feedback but employs \revision{a simple uniform random selection (\textbf{\rqrand})}.
Furthermore, we revisit the decision of using \gpt{4} as the analysis \llm by introducing an additional variant, \textbf{\rqstnl}, which employs \starcoder as the analysis \llm{}.

\parabf{Environment.} Our test-bed runs Ubuntu 20.04.5 LTS with 64-core CPUs, 256 GB RAM, and NVIDIA RTX A6000 GPUs.

\parabf{Fuzzing budget.}
Our default setting is to generate a total of 1000 tests for each optimization in 100 iterations.
In each iteration, \tech by default generates a batch of \batchvalue tests based on optimization triggering feedback from previous iterations. %
If the optimization was triggered in previous iterations, 
\tech picks three triggering inputs used as few-shot examples in the feedback-guided prompt (Section ~\ref{subsec:feedback}, Figure ~\ref{fig:prompt-feedback}) for the following iterations.
Otherwise, \tech uses the default few-shot prompt (Section ~\ref{subsec:test}, Figure ~\ref{fig:testgen-prompts}) to generate tests.

\subsection{Metrics}

Following prior work~\cite{nnsmith,titanfuzz,yarpgen,grayc,polyglot,tzer,donaldson2021test}, we use the \textit{number of detected bugs} as our metric, which essentially reflects the goal of fuzzing -- finding more bugs.
Meanwhile, the primary goal of our approach is to effectively test the optimizations within compilers.
As such, we also let the \textit{number of triggered optimizations} and the \textit{number of {(optimization-)}triggering tests} be our principal metrics.
Specifically, an optimization is deemed ``triggered'' if its corresponding optimization function (\S~\ref{sec:impl}) logs its presence during fuzzing.
Meanwhile, a test qualifies as a ``triggering test'' only if during its execution, any of the optimizations are triggered.
\revision{Given that optimization bugs can only manifest when the optimization is activated, similar to the concept of coverage, a higher number of \emph{triggering tests} correlates with an increased likelihood of bug discovery.}

\revision{To further show the effectiveness of every component, we also use \textit{code coverage}~\cite{titanfuzz,nnsmith,grayc,klees2018evaluating} as a metric.
Specifically, we report line coverage in the source languages where the optimizations are implemented: Python for \ptinductor, and C++ for both \tflite and \tfxla. Following previous work ~\cite{freefuzz,muffin,nablafuzz}, we measure line coverage using \CodeIn{Coverage.py}~\cite{coverage-py} for \python and GCOV~\cite{gcov} for \cpp.}

%% file: result.tex
\subsection{Comparison with Prior Work}
\label{subsec:baseline}

\begin{table*}[t]\centering
\caption{Comparison with baselines \revision{under the default setting}.}\label{tab:baseline}
\small
\begin{tabular}{l|r|r|r|r|r|rr}\toprule
& &\# Optim. &\# Triggered optim. &\# Triggering tests &\# Tests &Time/hour \\\midrule
&\tech &\multirow{4}{*}{61} &\textbf{41} &\textbf{21,469} &61,000 &41.1 \\
 \multirow{1}{*}{\pt} &\tech-Mini & &39 &1,737 &6,100 &4.2 \\
\multirow{1}{*}{\inductor} &\titanfuzz & &4 &5,519 &521,251 &76.6 \\
&\nnsmith & &5 &47 &12,084 &4.9 \\ \hline
\multirow{4}{*}{TF Lite} &\tech &\multirow{4}{*}{13} &\textbf{12} &2,801 &13,000 &18.1 \\
&\tech-Mini & &10 &305 &1,300 &1.1 \\
&\titanfuzz & &8 &571 &243,288 &59.0 \\
&\nnsmith & &7 &\textbf{4,666} &117,381 &6.8 \\ \hline
\multirow{4}{*}{TF-XLA} &\tech &\multirow{4}{*}{49} &20 &12,990 &49,000 &59.7 \\
&\tech-Mini & &19 &1,307 &4,900 &5.3 \\
&\titanfuzz & &\textbf{22} &45,762 &243,288 &63.2 \\
&\nnsmith & &16 &\textbf{117,006} &117,381 &6.0 \\
\bottomrule
\end{tabular}
\end{table*}

\begingroup
\begin{table}[t]\centering
\caption{\revision{Comparison with baselines over a 24-hour period.}}\label{tab:baseline-24}
\small
\begin{tabular}{l|r|r|r|r|r|rr}\toprule
& &\# Optim. &\# Triggered optim. &\# Triggering tests &\# Tests &Coverage \\\midrule
\multirow{1}{*}{\pt} &\tech &\multirow{3}{*}{61} &\textbf{41} &\textbf{12,127} &35,380 &\textbf{54,819} \\
\multirow{1}{*}{\inductor} &\titanfuzz & &4 &1,697 &167,521 &53,592 \\
&\nnsmith & &5 &233 &57,664 &49,910 \\
\midrule
\multirow{3}{*}{TF Lite} &\tech &\multirow{3}{*}{13} &\textbf{12} &3,369 &16,900 &{52,483} \\
&\titanfuzz & &7 &248 &126,364 &\textbf{55,606} \\
&\nnsmith & &8 &\textbf{19,747} &450,197 &28,108 \\
\midrule
\multirow{3}{*}{TF-XLA} &\tech &\multirow{3}{*}{49} &\textbf{19} &5,183 &19,600 &\textbf{66,224} \\
&\titanfuzz & &\textbf{19} &21,323 &115,351 &55,223 \\
&\nnsmith & &16 &\textbf{460,453} &460,970 &28,108 \\
\bottomrule
\end{tabular}
\end{table}
\endgroup

Table~\ref{tab:baseline} compares \tech against the baselines on the three target compilers \revision{under their default settings}.
Because \nnsmith has a shorter execution time than our default setting, for fair comparison, we also present results from \textbf{\tech-Mini}, which produces 100 tests for each optimization, as opposed to the default 1000 tests.
Notably, Column \textit{Time} in Table~\ref{tab:baseline} encompasses both the generation time of requirements/tests (including LLM invocations) and the test-execution time.

In terms of optimization triggering, we observe that \tech outperforms the baselines significantly in \ptinductor and \tflite.
Overall, among the tested compilers, \tech outperforms existing testers by up to 8.2x in terms of the number of triggered optimizations.
For example, out of the 61 optimizations in \ptinductor, \tech is able to trigger 41 optimizations, while the baseline approaches can trigger at most 5 optimizations, which is a subset of optimizations covered by \tech.
Regarding the time cost, \tech consumes less time than all other techniques except \nnsmith. Meanwhile, given the inferior performance of \nnsmith, \tech{}-Mini can still trigger more optimizations than \nnsmith using less time.

\tech outperforms all baselines on compilers except for \tfxla, with two fewer optimizations being triggered compared to \titanfuzz.
One possible reason is that the targeted optimizations in \tfxla are relatively simple per our optimization filtering for fair comparison with baselines (\S~\ref{sec:expsetup}).
Upon inspection, many of these optimizations represent common model patterns that are widely used in practice. 
Therefore, they can be effectively triggered by \titanfuzz since \titanfuzz leverages \llm{s} to generate human-like programs by resembling the distribution of training data.
Nevertheless, despite generating slightly fewer total tests compared to \titanfuzz, \tech demonstrates its own edge by triggering four unique optimizations which \titanfuzz cannot.
In addition, we note that \nnsmith has much more triggering tests than \tech and \titanfuzz over \tfxla.
This is largely due to the implementation choice of \nnsmith, which always outputs the model with redundant reshapes.
Thus, the vast majority of test inputs from \nnsmith can trigger the \CodeIn{IdentityReshapeRemoving} optimization (117,006/117,381).

\revision{Regarding unique optimizations triggered by each approach, the baselines trigger 7 optimizations for \ptinductor, while \tech covers these 7 and an additional 34 unique optimizations. For \tflite, the baselines trigger 9 optimizations, which are all covered by \tech, plus 3 more unique optimizations. For \tfxla, the baselines trigger 26 optimizations, including the 20 covered by \tech.}

\revision{
Additionally, Table~\ref{tab:baseline-24} compares \tech with the baselines over a 24-hour testing period, a common setting for fuzzing studies~\cite{klees2018evaluating}.
\tech performs the best on the number of triggered optimizations, substantially outperforming others on \ptinductor and \tflite.
In terms of code coverage, \tech covers more lines than the baselines in \ptinductor and \tfxla by up to 19.9\%.
For \tflite, \tech performs slightly worse than \titanfuzz (5.9\%).
This may be attributed to the limited number of optimizations (\emph{13}) in \tflite, which inherently restricts \tech's code coverage exploration, as \tech does not have much white-box information (i.e., optimization implementation) to guide the generation.
Meanwhile, please kindly note that code coverage is just a proxy indicator and does not always correlate strongly with bug finding abilities for complicated systems~\cite{inozemtseva2014coverage,su2021benchmarking}. 
Despite slightly lower code coverage on \tflite, \tech still performs better on the ultimate goal, bug finding (detailed in Section~\ref{subsec:bug}).
Overall, these results demonstrate the effectiveness of \tech in generating test cases to cover not only optimizations but also various compiler behaviors.
}

\subsection{Ablation Study}

\begin{table}[t]\centering
\caption{Impact of requirement description formats on \ptinductor.}
\label{tab:rqformat}
\small
\begin{tabular}{l|r|r}\toprule
&\# Triggered optim. (\% Total) &\# Triggering tests (\% Total) \\\midrule
\rqmix &\textbf{39} (\textbf{60.9\%}) &\textbf{1,113} (\textbf{17.4\%}) \\
\rqnl   &37 (57.8\%) &940   (14.7\%) \\
\rqcode &32 (50.0\%) &1,055 (16.5\%) \\
\rqimpl &32 (50.0\%) &638   (10.0\%) \\
\rqstnl &32 (50.0\%) &745   (11.6\%) \\
\bottomrule
\end{tabular}
\end{table}
{G}iven that \ptinductor has the highest number of optimizations, our ablation study is solely focused on \ptinductor.

\subsubsection{Requirement Generation \& Test Generation}
\label{subsec:rqstudy}
We first study the effectiveness of the requirement description and the multiple choices for the format (shown in Table~\ref{tab:rqformat}).
The goal of the requirement generation is to assist the generation \llm in producing more tests that can trigger additional optimizations within the compiler.
Thus, our main points of comparison are the number of triggered optimizations (Column ``\# Triggered optim.'') and the number of tests that can trigger the optimization (Column ``\# Triggered tests'').

\parabf{Effectiveness of requirement description.}
Compared with \rqimpl, which feeds the implementation source code directly with the generation \llm, all three variants that use requirements (\rqmix, \rqnl, and \rqcode) demonstrate superior performance in generating triggering tests.
Notably, our default setting \rqmix is able to generate \textit{1.74x} more triggering tests than directly using implementation code.
Besides, \rqmix can also trigger \textit{7} more optimizations than \rqimpl, emphasizing the importance of using requirement description.
This aligns with our statement in the Approach section that optimization source code is not the most effective guide for the generation \llm due to its redundant, unrelated information, and low-level format.

\parabf{Effectiveness of mixed format.}
As shown in Table~\ref{tab:rqformat}, \rqmix achieves the best number of triggered optimizations and triggering tests, underlining the effectiveness of combining \NL and \codeformat for requirement description.
Concurrently, while \rqnl triggers more optimizations than \rqcode, it results in fewer triggering tests.
This is because \NL usually contains additional information than \codeformat, ensuring vital triggering requirements are not missed during conversion from the implementation source code.
Conversely, it is more straightforward for the generation \llm to correlate requirements formatted in \codeformat with the respective test programs, leading to a higher number of triggering tests.

\parabf{Analysis \llm.}
When employing requirement descriptions generated by \starcoder, \rqstnl not only results in fewer triggered optimizations but also a reduced number of triggering tests compared to our default setting, which utilizes \gpt{4} to summarize the implementation source code.
This discrepancy is anticipated, given that \gpt{4} is recognized as the cutting-edge \llm in tasks related to code comprehension and natural language generation~\cite{bubeck2023sparks}.
Essentially, \gpt{4} exhibits a superior ability in translating intricate source code details into high-level input requirements compared to \starcoder.
This observation underscores our rationale for choosing \gpt{4} as the analytical \llm.
Interestingly, even \rqstnl generates a higher number of triggering tests than \rqimpl, which creates the input straight from the implementation source code.
Such a discovery confirms that a dual-model infrastructure might be better aligned for white-box compiler fuzzing than directly utilizing the implementation source code, emphasizing the value of having a distinct phase dedicated to requirement generation.

\subsubsection{Feedback Loop}
\label{subsec:fbstudy}

\begin{figure}[t]
    \centering
\includegraphics[keepaspectratio=true,width=0.95\columnwidth]{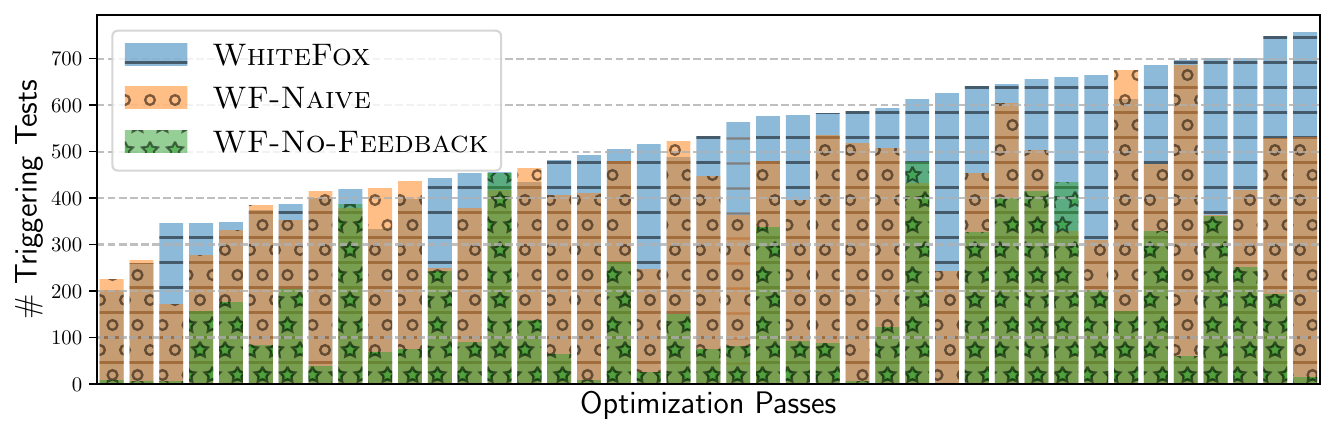}
    \caption{\revision{Impact of the feedback loop and Thompson Sampling on \ptinductor.}}
    \label{fig:feedback}
\end{figure}

\begin{table}[t]\centering
\caption{\revision{Statistics of the feedback loop on \ptinductor.}}\label{tab:feedback-stat}
\small
\begin{tabular}{l|r|rr}\toprule
&\# Triggering tests &Coverage \\\midrule
\tech &\textbf{21,469} &\textbf{55,857} \\
\rqrand &17,004 &54,602 \\
\rqnofb &8,152 &52,838 \\
\bottomrule
\end{tabular}
\end{table}

Next, we examine the effectiveness of our feedback loop and the Thompson Sampling algorithm.
The primary aim of the feedback loop is to enhance the likelihood of generating additional test inputs that activate already-triggered optimizations.
Therefore, in this ablation study, we focus on comparing the number of triggering tests.
Figure~\ref{fig:feedback} showcases a bar chart that details the number of triggering tests for each triggered optimization, spanning the range of variants explored in this ablation study.
\revision{Besides, we also collect the coverage results for these three variants, shown in Table~\ref{tab:feedback-stat}.}

\parabf{Effectiveness of feedback loop.}
\tech and \rqrand, incorporating the feedback loop, produce significantly more triggering tests than the variant without it (\rqnofb), with respectively \textit{2.6x} and \textit{2.1x} increases.
This improvement not only emphasizes the effectiveness of the feedback loop but indicates that its guidance is more attuned to triggering the target optimization than relying on few-shot examples for other different optimizations.
\revision{Furthermore, both approaches outperform \rqnofb in terms of code coverage, demonstrating that the feedback loop can guide \llm{s} to generate more diverse test cases.}

\parabf{Effectiveness of Thompson Sampling.}
In our default configuration, \tech leverages the Thompson Sampling algorithm for selecting triggering examples and achieves a remarkable \textit{1.3x} more triggering tests compared to \revision{\rqrand, which uses uniform sampling to select the examples}.
As shown in Figure~\ref{fig:feedback}, \tech outperforms the rest over 32 out of the 41 triggered optimizations in terms of the number of trigger tests. %
\revision{In the meanwhile, the code coverage of \tech is higher than \rqrand.}
Overall, the experimental results show the effectiveness of our MAB-based triggering example selection.

\subsection{Bug Finding}
\label{subsec:bug}

\begin{table}[!t]\centering
\caption{Overview of \tech{-detected} bugs.}\label{tab:bugfind}
\small
\begin{tabular}{l|r|rrr|rr}\toprule
&Total &Confirmed &New &Fixed &Won't fix\\\midrule
\pt &79 &76 &74 &68 &3  \\
\tflite &11 &8 &8 &0 &3  \\
\tfxla &11 &11 &10 &2 &0  \\\hline
Total &101 &95 &92 &70 &6  \\
\bottomrule
\end{tabular}
\end{table}

Table~\ref{tab:bugfind} presents the bug-finding results of \tech over \pt, \tfxla{, and \tflite}.
To date, \tech has detected \emph{\totalbug} bugs for them.
Of these, \newbug are confirmed as previously unknown, and \emph{\fixbug are already fixed}.
{For \pt, we also conducted bug detection in the latest version upon the request of the \pt team (will be discussed in \S~\ref{sec:discussion}). Thus, among the \pttotalbug identified bugs in the \ptinductor, 14 are found in its most recent version.
Remarkably, \emph{\highpriobug (12.7\%)} of the \pt bugs have been labeled with \textit{high priority}.
}

\revision{Of the 79 unique \tech-detected bugs, 68 are undetectable by the baselines. For \tfxla and \tflite, the baselines could find only 1 of the 22 bugs discovered by \tech.}

\subsubsection{Bug Analysis}
We next comprehensively analyze the \ptinductor bugs as it is the main source of \tech{-detected} bugs (\pttotalbug/\totalbug, \textit{\pttotalpercent\%}), and most of them have been fixed (\ptfixbug/\pttotalbug, \textit{\ptfixpercent\%}).
Out of these \pttotalbug bugs, only 11 (13.9\%) can be covered by the state-of-the-art, with 10 detectable by \titanfuzz and 3 by \nnsmith.

Regarding the \ptfixbug fixed bugs, we further explore the root cause of the bugs by inspecting their corresponding developer fixes.
Impressively, 47 (\textbf{69.1\%}) of the fixed bugs are repaired in the optimization code of \ptinductor.
This demonstrates the effectiveness of \tech for finding optimization bugs, which is the primary goal of our approach.
Specifically, only \textbf{3} of these 47 optimization bugs can be covered by \titanfuzz and \nnsmith, highlighting the significant edge of \tech in testing compiler optimizations.
One interesting observation is that certain optimizations appear to be more erroneous than the others;
however, such erroneous optimizations instead turn out to be harder to discover.
For example, \emph{\tech detects \emph{5} bugs in the optimization for the important attention modules~\cite{vaswani2017attention}, which are the fundamental building blocks to \llm{s}}.
The developer-crafted tests may seem surprising in their oversight of multiple critical bugs, but this is due to the challenge of creating precise model patterns to reveal deeply hidden issues. 
By exposing such critical bugs, \tech demonstrates the power of white-box fuzzing with \llm{s}.

\subsubsection{\revision{Bug Characteristics}}

\begin{table}[t]\centering
\caption{\revision{Characteristics of \tech-detected bugs}}
\label{tab:bug-char}
\small
\begin{tabular}{l|r|r|r|rr}\toprule
&Crash &Mis-compilation &Failed optim. &Incorrectly passed optim. (OOB)\\ \midrule
\ptinductor &6 &25 &41 &7 (3)\\
\tfxla &0 &4 &4 &3 (2)\\
\tflite &0 &8 &3 &0 (0)\\ \midrule
Total &6 &37 &48 &10 (5)\\
\bottomrule
\end{tabular}
\end{table}

\revision{
We further study the detailed characteristics of the \tech-detected bugs, which include \emph{crashes}, \emph{mis-compilations}, \emph{failed optimizations}, \emph{incorrectly passed optimizations}, and \emph{vulnerabilities}, shown in Table~\ref{tab:bug-char}. A \emph{mis-compilation} occurs when the optimized program returns different outputs than the non-optimized one. \emph{Failed optimizations} refer to cases where compilation with optimization fails, while it is valid without optimization. \emph{Incorrectly passed optimizations} occur when the optimization compiles invalid models successfully.
Regarding the \emph{vulnerabilities}, in addition to the 6 crash bugs that could be used for DoS attacks, there are another 5 out-of-bound read vulnerabilities detected within the \emph{incorrectly passed optimizations}.
}

\subsubsection{\revision{Won't Fix Bugs}}

\revision{
For the \emph{won't fix bugs}, in \ptinductor, one is due to the compiler not supporting quantized APIs, another is from undefined behavior in operators, and the third is because developers considered our input invalid, despite the optimization compiling the model and returning different results.
In \tflite, two bugs stem from its feature that doesn't guarantee input-output order, and another is the optimized output having different shapes, which is rare and not expected in both \ptinductor and \tfxla.
}

\begin{figure*}[t]
    \centering
\includegraphics[keepaspectratio=true,width=\textwidth]{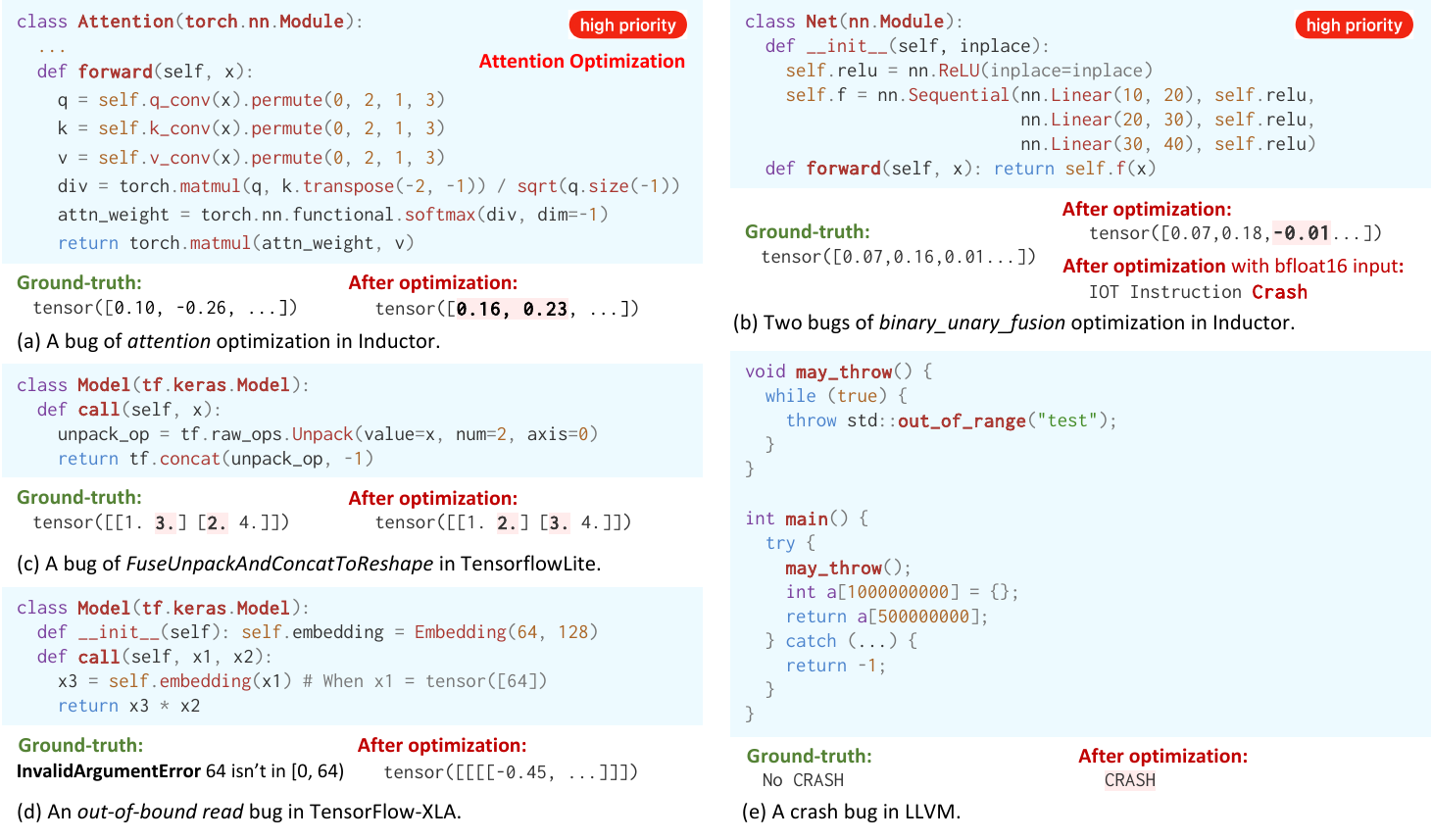}
    \caption{Example bugs detected by \tech.}
    \label{fig:bug}
\end{figure*}

\subsubsection{Bug Examples}
We demonstrate representative bugs detected by \tech and discuss their exploitation or security implications.
Figure~\ref{fig:bug}(a) illustrates a misoptimization of \ptinductor, manifested when compiling attention modules~\cite{vaswani2017attention}.
The faulty optimization through pattern matching identifies self-attentions and fuses their sub-operators into a compact and efficient implementation. {However, the optimized attention module, by rearranging the tensor layout to a channel-last format and rendering the last dimension non-contiguous, results in an accuracy issue.
This leads to incorrect outputs when compared to those from the unoptimized module.
Given the prevalence and impact of attention modules and \llm{s}, this bug is labeled with \emph{high-priority} and subsequently fixed.
The developers highlighted the importance of the issue, stating, \emph{``raising priority due to being an accuracy problem on an important operator''.}
}

{
Figure~\ref{fig:bug}(b) presents two bugs detected in \ptinductor for the \CodeIn{binary\_unary\_fusion} optimization, which fuses the \CodeIn{Linear} (\ie binary) and \CodeIn{ReLU} (\ie unary) operators into a compact and thus efficient operation.
However, after compilation, such exemplified module returns impermissible negative outputs since its final layer is \CodeIn{ReLU} whose output is always non-negative~\cite{agarap2018deep}.
Because such \CodeIn{Linear-ReLU} structures are incorporated in many fundamental architectures such as ResNet (Residual Network)~\cite{he2016identity},
this bug, uniquely found by \tech, is labeled as \emph{high-priority} and was fixed immediately after our report.
Furthermore, configuring the \CodeIn{ReLU} operation with \CodeIn{inplace=True} results in a crash of the optimized model for inputs of \CodeIn{bfloat16} data type, which can be leveraged for DoS attacks by requesting data in \CodeIn{bfloat16} format. 
Given the severity of this potential security issue, the developers promptly patched the vulnerability within two days.
}

Figure~\ref{fig:bug}(c) depicts a bug manifested in the \CodeIn{FuseUnpackAndConcatToReshape} optimization in \tflite.
Specifically, this optimization aims to streamline \CodeIn{unpack-concat} operation pairs into a single reshape operation, provided that the two operations are semantically inverse to each other.
In this optimization, the unpacked dimension should match the concatenated dimension in the original input of the \CodeIn{unpack-concat} operation pair.
The example listed in Figure~\ref{fig:bug}(c) violates the assumption and by theory cannot be simplified to a reshaping logic.
However, the \CodeIn{FuseUnpackAndConcatToReshape} function still erroneously transforms it into a wrong reshaping operation.
Notably, this bug is exclusively detected by \tech as it hinges on generating valid tests to trigger this particular optimization, which other techniques consistently struggle to accomplish.

{
Figure~\ref{fig:bug}(d) shows a \tfxla bug exclusively detected by \tech{}. The bug-triggering model contains an embedding layer with a vocabulary size of 64 tokens, followed by one multiplication operation.
When the first input to the model is 64, exceeding the embedding layer's maximum token index (63), the unoptimized model raises an \CodeIn{InvalidArgumentError} as expected.
However, the model optimized by XLA omits index validation, leading to an out-of-bounds read vulnerability.
Given the prevalence/impact of the important embedding layers, developers have swiftly addressed and fixed this issue after seeing our report.
}

%% file: discussion.tex
\subsection{Real-world Impact}
\label{sec:discussion}

Notably, we received the acknowledgment from the \pt team along with the request for integrating \tech into the development pipeline of \pt{} \inductor compiler.

\begin{tcolorbox}[width=0.95\textwidth, left=-2mm, right=-2mm, top=1mm, bottom=1mm, center]
\begin{quote}
\small
\textit{
``Thanks for your contributions to surfacing TorchInductor issues with Whitefox and sharing details. It will be great to figure out the next steps (for integration).'' --- \pt Team
}
\end{quote}
\end{tcolorbox}
\vspace{1mm}

Consequently, we further extend \tech to accommodate the most recent version of \ptinductor, incorporating support for an additional 38 newly introduced optimizations.
This underscores the distinct dynamic of \dl compilers, which diverge from traditional compilers due to the brisk pace of \dl model architecture evolution and the pressing need to optimize for nascent architectures.
For context, \ptinductor has experienced \emph{1,846 commits} in the last year alone.
Therefore, the principal focus of \tech on \dl compilers is driven by the necessity for an approach that can evolve in tandem with the rapid development of new optimizations.
As described in \S~\ref{subsec:bug}, \tech helped detect 14 new bugs for the newly introduced optimizations, all of which have been confirmed by the developers.
This underscores \tech's effectiveness and ability to adapt to evolving optimizations, showing the value and significance of incorporating \tech into the development workflow.

\subsection{Generality: Case Study on \llvm}

\begin{table*}[t]\centering
\caption{Comparison with baselines for \llvm optimizations}\label{tab:llvm}
\small
\begin{tabular}{l|r|r|r|r|r|rr}\toprule
& &\# Optim. &\# Triggered optim. &\# Triggering tests &\# Tests &Time/hour \\\midrule
\multirow{3}{*}{\llvm} &\tech &\multirow{3}{*}{52} &\textbf{26} &{25,322} &52,000 &30.9 \\
&\yarpgen & &4 &\textbf{76,171} &199,302	& 32.2 \\
&\grayc & &4 & 8,353 & 107,234 &30.6 \\
\bottomrule
\end{tabular}
\end{table*}

Although our main focus is on \dl compilers in this work, \tech is general to the compilers from various domains that contain pattern-based optimization pipelines.
To show the generality of \tech, we implemented a prototype of \tech for testing \llvm~\cite{llvm}, which is one of the most popular C/C++ compilers.
We tried our best to collect all \textit{middle-end} optimizations in \llvm since they are general to any architecture and are well-documented~\cite{llvmpass}.
In terms of baselines, we include \yarpgen~\cite{yarpgen}, a recent fuzzing tool to generate C/C++ test inputs with strategies to trigger different optimizations in compilers, and \grayc~\cite{grayc}, state-of-the-art grey-box fuzzer using coverage feedback to generate test programs in C.
\revision{Specifically, the \llvm version under test is LLVM-18-20230818-nightly, and the experimental environment matches the setup described in \S~\ref{sec:expsetup}.}

\parabf{Optimization trigger.} Table~\ref{tab:llvm} presents the optimization triggering results for \llvm optimizations against the baselines.
Similar to the results on \dl compilers, we observe that \tech can trigger \emph{6.5x} more optimizations than baselines while incurring less time cost.
This demonstrates the effectiveness and potential of \tech on different compilers.

\parabf{Bug detection.} \tech detects 6 bugs for \llvm, with 2 confirmed as previously unknown, 3 pending, and 1 won't fix.
Figure~\ref{fig:bug}(e) presents a confirmed \llvm bug, 
which is only revealed 
when a test program references a huge array through a large enough index, 
crashing the \llvm post-optimization.
Attackers can exploit this vulnerability for DoS by crafting specific input programs to crash Just-in-Time-enabled systems that use this optimization.

\subsection{\revision{Limitations and Future Work}}

\revision{
While this paper focuses on compiler optimization, \tech could be potentially adapted for white-box fuzzing of other compiler code and even other complex, real-world software systems.
For example, for regression bugs, \tech can be deployed by setting the changed branches as targets and letting \llm{s} analyze their triggering conditions.
However, one limitation or challenge is that, for optimization, the high-level system inputs usually have a relatively clear mapping to the low-level optimization implementation. In contrast, for arbitrary functions, this mapping may be unclear. To mitigate this, one possibility is to further leverage \llm{s} to infer such mappings (\eg leveraging auxiliary information, including documentation) along with the triggering conditions.
}

\revision{
Another possible future direction is to use traditional fuzzing approaches as external tools for efficient test generation.
Given the higher computational cost of invoking smaller \llm{s} compared to traditional techniques, this strategy could improve performance. For instance, we could utilize \tech to summarize optimization-triggering generation or mutation rules, which could then guide input generation for a traditional fuzzing framework such as \nnsmith~\cite{nnsmith}.
}

%% file: main.bbl

\begin{thebibliography}{99}

%%% the Web address.

\ifx \showCODEN    \undefined \def \showCODEN     #1{\unskip}     \fi
\ifx \showDOI      \undefined \def \showDOI       #1{#1}\fi
\ifx \showISBNx    \undefined \def \showISBNx     #1{\unskip}     \fi
\ifx \showISBNxiii \undefined \def \showISBNxiii  #1{\unskip}     \fi
\ifx \showISSN     \undefined \def \showISSN      #1{\unskip}     \fi
\ifx \showLCCN     \undefined \def \showLCCN      #1{\unskip}     \fi
\ifx \shownote     \undefined \def \shownote      #1{#1}          \fi
\ifx \showarticletitle \undefined \def \showarticletitle #1{#1}   \fi
\ifx \showURL      \undefined \def \showURL       {\relax}        \fi
% invisible to TeX
\providecommand\bibfield[2]{#2}
\providecommand\bibinfo[2]{#2}
\providecommand\natexlab[1]{#1}
\providecommand\showeprint[2][]{arXiv:#2}

\bibitem[ube(2021)]%
        {uberkill}
 \bibinfo{year}{2021}\natexlab{}.
\newblock \bibinfo{title}{{News}}.
\newblock
\newblock
\newblock
\shownote{https://www.vice.com/en\_us/article/9kga85/uber-is-giving-up-on-self-driving-cars-in-california-after-deadly-crash}.


\bibitem[cov(2022)]%
        {coverage-py}
 \bibinfo{year}{2022}\natexlab{}.
\newblock \bibinfo{title}{{Coverage.py}}.
\newblock \bibinfo{howpublished}{\url{https://github.com/nedbat/coveragepy}}.
\newblock


\bibitem[gco(2022)]%
        {gcov}
 \bibinfo{year}{2022}\natexlab{}.
\newblock \bibinfo{title}{{GCOV}}.
\newblock \bibinfo{howpublished}{\url{https://gcc.gnu.org/onlinedocs/gcc/Gcov.html}}.
\newblock


\bibitem[Agarap(2018)]%
        {agarap2018deep}
\bibfield{author}{\bibinfo{person}{Abien~Fred Agarap}.} \bibinfo{year}{2018}\natexlab{}.
\newblock \showarticletitle{Deep learning using rectified linear units (relu)}.
\newblock \bibinfo{journal}{\emph{arXiv preprint arXiv:1803.08375}} (\bibinfo{year}{2018}).
\newblock


\bibitem[Brown et~al\mbox{.}(2020)]%
        {fewshot}
\bibfield{author}{\bibinfo{person}{Tom Brown}, \bibinfo{person}{Benjamin Mann}, \bibinfo{person}{Nick Ryder}, \bibinfo{person}{Melanie Subbiah}, \bibinfo{person}{Jared~D Kaplan}, \bibinfo{person}{Prafulla Dhariwal}, \bibinfo{person}{Arvind Neelakantan}, \bibinfo{person}{Pranav Shyam}, \bibinfo{person}{Girish Sastry}, \bibinfo{person}{Amanda Askell}, {et~al\mbox{.}}} \bibinfo{year}{2020}\natexlab{}.
\newblock \showarticletitle{Language models are few-shot learners}.
\newblock \bibinfo{journal}{\emph{Advances in neural information processing systems}}  \bibinfo{volume}{33} (\bibinfo{year}{2020}), \bibinfo{pages}{1877--1901}.
\newblock


\bibitem[Bubeck et~al\mbox{.}(2023)]%
        {bubeck2023sparks}
\bibfield{author}{\bibinfo{person}{S{\'e}bastien Bubeck}, \bibinfo{person}{Varun Chandrasekaran}, \bibinfo{person}{Ronen Eldan}, \bibinfo{person}{Johannes Gehrke}, \bibinfo{person}{Eric Horvitz}, \bibinfo{person}{Ece Kamar}, \bibinfo{person}{Peter Lee}, \bibinfo{person}{Yin~Tat Lee}, \bibinfo{person}{Yuanzhi Li}, \bibinfo{person}{Scott Lundberg}, {et~al\mbox{.}}} \bibinfo{year}{2023}\natexlab{}.
\newblock \showarticletitle{Sparks of artificial general intelligence: Early experiments with gpt-4}.
\newblock \bibinfo{journal}{\emph{arXiv preprint arXiv:2303.12712}} (\bibinfo{year}{2023}).
\newblock


\bibitem[Chen et~al\mbox{.}(2020)]%
        {chen2020survey}
\bibfield{author}{\bibinfo{person}{Junjie Chen}, \bibinfo{person}{Jibesh Patra}, \bibinfo{person}{Michael Pradel}, \bibinfo{person}{Yingfei Xiong}, \bibinfo{person}{Hongyu Zhang}, \bibinfo{person}{Dan Hao}, {and} \bibinfo{person}{Lu Zhang}.} \bibinfo{year}{2020}\natexlab{}.
\newblock \showarticletitle{A survey of compiler testing}.
\newblock \bibinfo{journal}{\emph{ACM Computing Surveys (CSUR)}} \bibinfo{volume}{53}, \bibinfo{number}{1} (\bibinfo{year}{2020}), \bibinfo{pages}{1--36}.
\newblock


\bibitem[Chen et~al\mbox{.}(2021a)]%
        {codex}
\bibfield{author}{\bibinfo{person}{Mark Chen}, \bibinfo{person}{Jerry Tworek}, \bibinfo{person}{Heewoo Jun}, \bibinfo{person}{Qiming Yuan}, \bibinfo{person}{Henrique Ponde de~Oliveira Pinto}, \bibinfo{person}{Jared Kaplan}, \bibinfo{person}{Harri Edwards}, \bibinfo{person}{Yuri Burda}, \bibinfo{person}{Nicholas Joseph}, \bibinfo{person}{Greg Brockman}, {et~al\mbox{.}}} \bibinfo{year}{2021}\natexlab{a}.
\newblock \showarticletitle{Evaluating large language models trained on code}.
\newblock \bibinfo{journal}{\emph{arXiv preprint arXiv:2107.03374}} (\bibinfo{year}{2021}).
\newblock


\bibitem[Chen et~al\mbox{.}(2018)]%
        {chen2018tvm}
\bibfield{author}{\bibinfo{person}{Tianqi Chen}, \bibinfo{person}{Thierry Moreau}, \bibinfo{person}{Ziheng Jiang}, \bibinfo{person}{Lianmin Zheng}, \bibinfo{person}{Eddie Yan}, \bibinfo{person}{Haichen Shen}, \bibinfo{person}{Meghan Cowan}, \bibinfo{person}{Leyuan Wang}, \bibinfo{person}{Yuwei Hu}, \bibinfo{person}{Luis Ceze}, {et~al\mbox{.}}} \bibinfo{year}{2018}\natexlab{}.
\newblock \showarticletitle{$\{$TVM$\}$: An automated $\{$End-to-End$\}$ optimizing compiler for deep learning}. In \bibinfo{booktitle}{\emph{13th USENIX Symposium on Operating Systems Design and Implementation (OSDI 18)}}. \bibinfo{pages}{578--594}.
\newblock


\bibitem[Chen et~al\mbox{.}(2024)]%
        {chen2024chatunitest}
\bibfield{author}{\bibinfo{person}{Yinghao Chen}, \bibinfo{person}{Zehao Hu}, \bibinfo{person}{Chen Zhi}, \bibinfo{person}{Junxiao Han}, \bibinfo{person}{Shuiguang Deng}, {and} \bibinfo{person}{Jianwei Yin}.} \bibinfo{year}{2024}\natexlab{}.
\newblock \showarticletitle{ChatUniTest: A Framework for LLM-Based Test Generation}. In \bibinfo{booktitle}{\emph{Companion Proceedings of the 32nd ACM International Conference on the Foundations of Software Engineering}}. \bibinfo{pages}{572--576}.
\newblock


\bibitem[Chen et~al\mbox{.}(2021b)]%
        {polyglot}
\bibfield{author}{\bibinfo{person}{Yongheng Chen}, \bibinfo{person}{Rui Zhong}, \bibinfo{person}{Hong Hu}, \bibinfo{person}{Hangfan Zhang}, \bibinfo{person}{Yupeng Yang}, \bibinfo{person}{Dinghao Wu}, {and} \bibinfo{person}{Wenke Lee}.} \bibinfo{year}{2021}\natexlab{b}.
\newblock \showarticletitle{One engine to fuzz’em all: Generic language processor testing with semantic validation}. In \bibinfo{booktitle}{\emph{2021 IEEE Symposium on Security and Privacy (SP)}}. IEEE, \bibinfo{pages}{642--658}.
\newblock


\bibitem[Cho et~al\mbox{.}(2019)]%
        {cho2019intriguer}
\bibfield{author}{\bibinfo{person}{Mingi Cho}, \bibinfo{person}{Seoyoung Kim}, {and} \bibinfo{person}{Taekyoung Kwon}.} \bibinfo{year}{2019}\natexlab{}.
\newblock \showarticletitle{Intriguer: Field-level constraint solving for hybrid fuzzing}. In \bibinfo{booktitle}{\emph{Proceedings of the 2019 ACM SIGSAC Conference on Computer and Communications Security}}. \bibinfo{pages}{515--530}.
\newblock


\bibitem[Choi et~al\mbox{.}(2019)]%
        {choi2019grey}
\bibfield{author}{\bibinfo{person}{Jaeseung Choi}, \bibinfo{person}{Joonun Jang}, \bibinfo{person}{Choongwoo Han}, {and} \bibinfo{person}{Sang~Kil Cha}.} \bibinfo{year}{2019}\natexlab{}.
\newblock \showarticletitle{Grey-box concolic testing on binary code}. In \bibinfo{booktitle}{\emph{2019 IEEE/ACM 41st International Conference on Software Engineering (ICSE)}}. IEEE, \bibinfo{pages}{736--747}.
\newblock


\bibitem[Chowdhery et~al\mbox{.}(2022)]%
        {palm}
\bibfield{author}{\bibinfo{person}{Aakanksha Chowdhery}, \bibinfo{person}{Sharan Narang}, \bibinfo{person}{Jacob Devlin}, \bibinfo{person}{Maarten Bosma}, \bibinfo{person}{Gaurav Mishra}, \bibinfo{person}{Adam Roberts}, \bibinfo{person}{Paul Barham}, \bibinfo{person}{Hyung~Won Chung}, \bibinfo{person}{Charles Sutton}, \bibinfo{person}{Sebastian Gehrmann}, {et~al\mbox{.}}} \bibinfo{year}{2022}\natexlab{}.
\newblock \showarticletitle{Palm: Scaling language modeling with pathways}.
\newblock \bibinfo{journal}{\emph{arXiv preprint arXiv:2204.02311}} (\bibinfo{year}{2022}).
\newblock


\bibitem[Christou et~al\mbox{.}(2023)]%
        {christou2023ivysyn}
\bibfield{author}{\bibinfo{person}{Neophytos Christou}, \bibinfo{person}{Di Jin}, \bibinfo{person}{Vaggelis Atlidakis}, \bibinfo{person}{Baishakhi Ray}, {and} \bibinfo{person}{Vasileios~P Kemerlis}.} \bibinfo{year}{2023}\natexlab{}.
\newblock \showarticletitle{$\{$IvySyn$\}$: Automated Vulnerability Discovery in Deep Learning Frameworks}. In \bibinfo{booktitle}{\emph{32nd USENIX Security Symposium (USENIX Security 23)}}. \bibinfo{pages}{2383--2400}.
\newblock


\bibitem[Dakhel et~al\mbox{.}(2024)]%
        {dakhel2024effective}
\bibfield{author}{\bibinfo{person}{Arghavan~Moradi Dakhel}, \bibinfo{person}{Amin Nikanjam}, \bibinfo{person}{Vahid Majdinasab}, \bibinfo{person}{Foutse Khomh}, {and} \bibinfo{person}{Michel~C Desmarais}.} \bibinfo{year}{2024}\natexlab{}.
\newblock \showarticletitle{Effective test generation using pre-trained large language models and mutation testing}.
\newblock \bibinfo{journal}{\emph{Information and Software Technology}}  \bibinfo{volume}{171} (\bibinfo{year}{2024}), \bibinfo{pages}{107468}.
\newblock


\bibitem[Deng et~al\mbox{.}(2023a)]%
        {titanfuzz}
\bibfield{author}{\bibinfo{person}{Yinlin Deng}, \bibinfo{person}{Chunqiu~Steven Xia}, \bibinfo{person}{Haoran Peng}, \bibinfo{person}{Chenyuan Yang}, {and} \bibinfo{person}{Lingming Zhang}.} \bibinfo{year}{2023}\natexlab{a}.
\newblock \showarticletitle{Large Language Models are Zero-Shot Fuzzers: Fuzzing Deep-Learning Libraries via Large Language Models}. In \bibinfo{booktitle}{\emph{Proceedings of the 32nd ACM SIGSOFT International Symposium on Software Testing and Analysis}} \emph{(\bibinfo{series}{ISSTA 2023})}.
\newblock


\bibitem[Deng et~al\mbox{.}(2023b)]%
        {deng2023large}
\bibfield{author}{\bibinfo{person}{Yinlin Deng}, \bibinfo{person}{Chunqiu~Steven Xia}, \bibinfo{person}{Chenyuan Yang}, \bibinfo{person}{Shizhuo~Dylan Zhang}, \bibinfo{person}{Shujing Yang}, {and} \bibinfo{person}{Lingming Zhang}.} \bibinfo{year}{2023}\natexlab{b}.
\newblock \showarticletitle{Large language models are edge-case fuzzers: Testing deep learning libraries via fuzzgpt}.
\newblock \bibinfo{journal}{\emph{arXiv preprint arXiv:2304.02014}} (\bibinfo{year}{2023}).
\newblock


\bibitem[Donaldson et~al\mbox{.}(2017)]%
        {donaldson2017automated}
\bibfield{author}{\bibinfo{person}{Alastair~F Donaldson}, \bibinfo{person}{Hugues Evrard}, \bibinfo{person}{Andrei Lascu}, {and} \bibinfo{person}{Paul Thomson}.} \bibinfo{year}{2017}\natexlab{}.
\newblock \showarticletitle{Automated testing of graphics shader compilers}.
\newblock \bibinfo{journal}{\emph{Proceedings of the ACM on Programming Languages}} \bibinfo{volume}{1}, \bibinfo{number}{OOPSLA} (\bibinfo{year}{2017}), \bibinfo{pages}{1--29}.
\newblock


\bibitem[Donaldson et~al\mbox{.}(2021)]%
        {donaldson2021test}
\bibfield{author}{\bibinfo{person}{Alastair~F Donaldson}, \bibinfo{person}{Paul Thomson}, \bibinfo{person}{Vasyl Teliman}, \bibinfo{person}{Stefano Milizia}, \bibinfo{person}{Andr{\'e}~Perez Maselco}, {and} \bibinfo{person}{Antoni Karpi{\'n}ski}.} \bibinfo{year}{2021}\natexlab{}.
\newblock \showarticletitle{Test-case reduction and deduplication almost for free with transformation-based compiler testing}. In \bibinfo{booktitle}{\emph{Proceedings of the 42nd ACM SIGPLAN International Conference on Programming Language Design and Implementation}}. \bibinfo{pages}{1017--1032}.
\newblock


\bibitem[Even-Mendoza et~al\mbox{.}(2023)]%
        {grayc}
\bibfield{author}{\bibinfo{person}{Karine Even-Mendoza}, \bibinfo{person}{Arindam Sharma}, \bibinfo{person}{Alastair~F Donaldson}, {and} \bibinfo{person}{Cristian Cadar}.} \bibinfo{year}{2023}\natexlab{}.
\newblock \showarticletitle{GrayC: Greybox Fuzzing of Compilers and Analysers for C}.
\newblock  (\bibinfo{year}{2023}).
\newblock


\bibitem[Feng et~al\mbox{.}(2020)]%
        {feng2020codebert}
\bibfield{author}{\bibinfo{person}{Zhangyin Feng}, \bibinfo{person}{Daya Guo}, \bibinfo{person}{Duyu Tang}, \bibinfo{person}{Nan Duan}, \bibinfo{person}{Xiaocheng Feng}, \bibinfo{person}{Ming Gong}, \bibinfo{person}{Linjun Shou}, \bibinfo{person}{Bing Qin}, \bibinfo{person}{Ting Liu}, \bibinfo{person}{Daxin Jiang}, {et~al\mbox{.}}} \bibinfo{year}{2020}\natexlab{}.
\newblock \showarticletitle{Codebert: A pre-trained model for programming and natural languages}.
\newblock \bibinfo{journal}{\emph{arXiv preprint arXiv:2002.08155}} (\bibinfo{year}{2020}).
\newblock


\bibitem[Fraser and Arcuri(2011)]%
        {fraser2011evosuite}
\bibfield{author}{\bibinfo{person}{Gordon Fraser} {and} \bibinfo{person}{Andrea Arcuri}.} \bibinfo{year}{2011}\natexlab{}.
\newblock \showarticletitle{Evosuite: automatic test suite generation for object-oriented software}. In \bibinfo{booktitle}{\emph{Proceedings of the 19th ACM SIGSOFT symposium and the 13th European conference on Foundations of software engineering}}. \bibinfo{pages}{416--419}.
\newblock


\bibitem[GCC(2023)]%
        {gcc}
GCC \bibinfo{year}{2023}\natexlab{}.
\newblock \bibinfo{title}{GCC}.
\newblock
\newblock
\newblock
\shownote{\url{https://gcc.gnu.org/}}.


\bibitem[Godefroid et~al\mbox{.}(2005)]%
        {dart}
\bibfield{author}{\bibinfo{person}{Patrice Godefroid}, \bibinfo{person}{Nils Klarlund}, {and} \bibinfo{person}{Koushik Sen}.} \bibinfo{year}{2005}\natexlab{}.
\newblock \showarticletitle{DART: Directed Automated Random Testing}. In \bibinfo{booktitle}{\emph{Proceedings of the 2005 ACM SIGPLAN Conference on Programming Language Design and Implementation}} (Chicago, IL, USA) \emph{(\bibinfo{series}{PLDI '05})}. \bibinfo{publisher}{Association for Computing Machinery}, \bibinfo{address}{New York, NY, USA}, \bibinfo{pages}{213–223}.
\newblock
\showISBNx{1595930566}
\urldef\tempurl%
\url{https://doi.org/10.1145/1065010.1065036}
\showDOI{\tempurl}


\bibitem[Gopinath et~al\mbox{.}(2020)]%
        {gopinath2020fuzzing}
\bibfield{author}{\bibinfo{person}{Rahul Gopinath}, \bibinfo{person}{Bachir Bendrissou}, \bibinfo{person}{Bj{\"o}rn Mathis}, {and} \bibinfo{person}{Andreas Zeller}.} \bibinfo{year}{2020}\natexlab{}.
\newblock \showarticletitle{Fuzzing with fast failure feedback}.
\newblock \bibinfo{journal}{\emph{arXiv preprint arXiv:2012.13516}} (\bibinfo{year}{2020}).
\newblock


\bibitem[Gu et~al\mbox{.}(2022)]%
        {muffin}
\bibfield{author}{\bibinfo{person}{J. Gu}, \bibinfo{person}{X. Luo}, \bibinfo{person}{Y. Zhou}, {and} \bibinfo{person}{X. Wang}.} \bibinfo{year}{2022}\natexlab{}.
\newblock \showarticletitle{Muffin: Testing Deep Learning Libraries via Neural Architecture Fuzzing}. In \bibinfo{booktitle}{\emph{2022 IEEE/ACM 44th International Conference on Software Engineering (ICSE)}}. \bibinfo{publisher}{IEEE Computer Society}, \bibinfo{address}{Los Alamitos, CA, USA}, \bibinfo{pages}{1418--1430}.
\newblock
\urldef\tempurl%
\url{https://doi.org/10.1145/3510003.3510092}
\showDOI{\tempurl}


\bibitem[Guo et~al\mbox{.}(2020)]%
        {audee}
\bibfield{author}{\bibinfo{person}{Qianyu Guo}, \bibinfo{person}{Xiaofei Xie}, \bibinfo{person}{Yi Li}, \bibinfo{person}{Xiaoyu Zhang}, \bibinfo{person}{Yang Liu}, \bibinfo{person}{Xiaohong Li}, {and} \bibinfo{person}{Chao Shen}.} \bibinfo{year}{2020}\natexlab{}.
\newblock \showarticletitle{Audee: Automated testing for deep learning frameworks}. In \bibinfo{booktitle}{\emph{2020 35th IEEE/ACM International Conference on Automated Software Engineering (ASE)}}. IEEE, \bibinfo{pages}{486--498}.
\newblock


\bibitem[He et~al\mbox{.}(2016)]%
        {he2016identity}
\bibfield{author}{\bibinfo{person}{Kaiming He}, \bibinfo{person}{Xiangyu Zhang}, \bibinfo{person}{Shaoqing Ren}, {and} \bibinfo{person}{Jian Sun}.} \bibinfo{year}{2016}\natexlab{}.
\newblock \showarticletitle{Identity mappings in deep residual networks}. In \bibinfo{booktitle}{\emph{Computer Vision--ECCV 2016: 14th European Conference, Amsterdam, The Netherlands, October 11--14, 2016, Proceedings, Part IV 14}}. Springer, \bibinfo{pages}{630--645}.
\newblock


\bibitem[Holler et~al\mbox{.}(2012)]%
        {holler2012fuzzing}
\bibfield{author}{\bibinfo{person}{Christian Holler}, \bibinfo{person}{Kim Herzig}, {and} \bibinfo{person}{Andreas Zeller}.} \bibinfo{year}{2012}\natexlab{}.
\newblock \showarticletitle{Fuzzing with code fragments}. In \bibinfo{booktitle}{\emph{21st {USENIX} Security Symposium ({USENIX} Security 12)}}. \bibinfo{pages}{445--458}.
\newblock


\bibitem[Huang et~al\mbox{.}(2020)]%
        {huang2020pangolin}
\bibfield{author}{\bibinfo{person}{Heqing Huang}, \bibinfo{person}{Peisen Yao}, \bibinfo{person}{Rongxin Wu}, \bibinfo{person}{Qingkai Shi}, {and} \bibinfo{person}{Charles Zhang}.} \bibinfo{year}{2020}\natexlab{}.
\newblock \showarticletitle{Pangolin: Incremental hybrid fuzzing with polyhedral path abstraction}. In \bibinfo{booktitle}{\emph{2020 IEEE Symposium on Security and Privacy (SP)}}. IEEE, \bibinfo{pages}{1613--1627}.
\newblock


\bibitem[Huang et~al\mbox{.}(2024)]%
        {huang2024large}
\bibfield{author}{\bibinfo{person}{Linghan Huang}, \bibinfo{person}{Peizhou Zhao}, \bibinfo{person}{Huaming Chen}, {and} \bibinfo{person}{Lei Ma}.} \bibinfo{year}{2024}\natexlab{}.
\newblock \showarticletitle{Large language models based fuzzing techniques: A survey}.
\newblock \bibinfo{journal}{\emph{arXiv preprint arXiv:2402.00350}} (\bibinfo{year}{2024}).
\newblock


\bibitem[HuggingFace(2023)]%
        {HuggingFaceWebPage}
HuggingFace \bibinfo{year}{2023}\natexlab{}.
\newblock \bibinfo{title}{Hugging Face}.
\newblock
\newblock
\newblock
\shownote{\url{https://huggingface.co}}.


\bibitem[Inozemtseva and Holmes(2014)]%
        {inozemtseva2014coverage}
\bibfield{author}{\bibinfo{person}{Laura Inozemtseva} {and} \bibinfo{person}{Reid Holmes}.} \bibinfo{year}{2014}\natexlab{}.
\newblock \showarticletitle{Coverage is not strongly correlated with test suite effectiveness}. In \bibinfo{booktitle}{\emph{Proceedings of the 36th international conference on software engineering}}. \bibinfo{pages}{435--445}.
\newblock


\bibitem[Jiang et~al\mbox{.}(2023)]%
        {jiang2023evaluating}
\bibfield{author}{\bibinfo{person}{Ling Jiang}, \bibinfo{person}{Hengchen Yuan}, \bibinfo{person}{Mingyuan Wu}, \bibinfo{person}{Lingming Zhang}, {and} \bibinfo{person}{Yuqun Zhang}.} \bibinfo{year}{2023}\natexlab{}.
\newblock \showarticletitle{Evaluating and improving hybrid fuzzing}. In \bibinfo{booktitle}{\emph{2023 IEEE/ACM 45th International Conference on Software Engineering (ICSE)}}. IEEE, \bibinfo{pages}{410--422}.
\newblock


\bibitem[Kim et~al\mbox{.}(2020)]%
        {kim2020hfl}
\bibfield{author}{\bibinfo{person}{Kyungtae Kim}, \bibinfo{person}{Dae~R Jeong}, \bibinfo{person}{Chung~Hwan Kim}, \bibinfo{person}{Yeongjin Jang}, \bibinfo{person}{Insik Shin}, {and} \bibinfo{person}{Byoungyoung Lee}.} \bibinfo{year}{2020}\natexlab{}.
\newblock \showarticletitle{HFL: Hybrid Fuzzing on the Linux Kernel.}. In \bibinfo{booktitle}{\emph{NDSS}}.
\newblock


\bibitem[King(1976)]%
        {symbolic}
\bibfield{author}{\bibinfo{person}{James~C King}.} \bibinfo{year}{1976}\natexlab{}.
\newblock \showarticletitle{Symbolic execution and program testing}.
\newblock \bibinfo{journal}{\emph{Commun. ACM}} \bibinfo{volume}{19}, \bibinfo{number}{7} (\bibinfo{year}{1976}), \bibinfo{pages}{385--394}.
\newblock


\bibitem[Klees et~al\mbox{.}(2018)]%
        {klees2018evaluating}
\bibfield{author}{\bibinfo{person}{George Klees}, \bibinfo{person}{Andrew Ruef}, \bibinfo{person}{Benji Cooper}, \bibinfo{person}{Shiyi Wei}, {and} \bibinfo{person}{Michael Hicks}.} \bibinfo{year}{2018}\natexlab{}.
\newblock \showarticletitle{Evaluating fuzz testing}. In \bibinfo{booktitle}{\emph{Proceedings of the 2018 ACM SIGSAC conference on computer and communications security}}. \bibinfo{pages}{2123--2138}.
\newblock


\bibitem[Lattner and Adve(2004)]%
        {llvm}
\bibfield{author}{\bibinfo{person}{Chris Lattner} {and} \bibinfo{person}{Vikram Adve}.} \bibinfo{year}{2004}\natexlab{}.
\newblock \showarticletitle{LLVM: A compilation framework for lifelong program analysis \& transformation}. In \bibinfo{booktitle}{\emph{International symposium on code generation and optimization, 2004. CGO 2004.}} IEEE, \bibinfo{pages}{75--86}.
\newblock


\bibitem[Le et~al\mbox{.}(2014)]%
        {le2014compiler}
\bibfield{author}{\bibinfo{person}{Vu Le}, \bibinfo{person}{Mehrdad Afshari}, {and} \bibinfo{person}{Zhendong Su}.} \bibinfo{year}{2014}\natexlab{}.
\newblock \showarticletitle{Compiler validation via equivalence modulo inputs}.
\newblock \bibinfo{journal}{\emph{ACM Sigplan Notices}} \bibinfo{volume}{49}, \bibinfo{number}{6} (\bibinfo{year}{2014}), \bibinfo{pages}{216--226}.
\newblock


\bibitem[Le et~al\mbox{.}(2015)]%
        {le2015finding}
\bibfield{author}{\bibinfo{person}{Vu Le}, \bibinfo{person}{Chengnian Sun}, {and} \bibinfo{person}{Zhendong Su}.} \bibinfo{year}{2015}\natexlab{}.
\newblock \showarticletitle{Finding deep compiler bugs via guided stochastic program mutation}.
\newblock \bibinfo{journal}{\emph{ACM SIGPLAN Notices}} \bibinfo{volume}{50}, \bibinfo{number}{10} (\bibinfo{year}{2015}), \bibinfo{pages}{386--399}.
\newblock


\bibitem[Lemieux et~al\mbox{.}(2023)]%
        {codamosa}
\bibfield{author}{\bibinfo{person}{Caroline Lemieux}, \bibinfo{person}{Jeevana~Priya Inala}, \bibinfo{person}{Shuvendu~K Lahiri}, {and} \bibinfo{person}{Siddhartha Sen}.} \bibinfo{year}{2023}\natexlab{}.
\newblock \showarticletitle{CODAMOSA: Escaping coverage plateaus in test generation with pre-trained large language models}. In \bibinfo{booktitle}{\emph{International conference on software engineering (ICSE)}}.
\newblock


\bibitem[Li et~al\mbox{.}(2023a)]%
        {starcoder}
\bibfield{author}{\bibinfo{person}{Raymond Li}, \bibinfo{person}{Loubna~Ben Allal}, \bibinfo{person}{Yangtian Zi}, \bibinfo{person}{Niklas Muennighoff}, \bibinfo{person}{Denis Kocetkov}, \bibinfo{person}{Chenghao Mou}, \bibinfo{person}{Marc Marone}, \bibinfo{person}{Christopher Akiki}, \bibinfo{person}{Jia Li}, \bibinfo{person}{Jenny Chim}, {et~al\mbox{.}}} \bibinfo{year}{2023}\natexlab{a}.
\newblock \showarticletitle{StarCoder: may the source be with you!}
\newblock \bibinfo{journal}{\emph{arXiv preprint arXiv:2305.06161}} (\bibinfo{year}{2023}).
\newblock


\bibitem[Li et~al\mbox{.}(2023b)]%
        {li2023pyrtfuzz}
\bibfield{author}{\bibinfo{person}{Wen Li}, \bibinfo{person}{Haoran Yang}, \bibinfo{person}{Xiapu Luo}, \bibinfo{person}{Long Cheng}, {and} \bibinfo{person}{Haipeng Cai}.} \bibinfo{year}{2023}\natexlab{b}.
\newblock \showarticletitle{PyRTFuzz: Detecting Bugs in Python Runtimes via Two-Level Collaborative Fuzzing}. In \bibinfo{booktitle}{\emph{Proceedings of the 2023 ACM SIGSAC Conference on Computer and Communications Security}}. \bibinfo{pages}{1645--1659}.
\newblock


\bibitem[libFuzzer(2023)]%
        {libfuzzer}
libFuzzer \bibinfo{year}{2023}\natexlab{}.
\newblock \bibinfo{title}{libFuzzer – a library for coverage-guided fuzz testing.}
\newblock
\newblock
\newblock
\shownote{\url{https://llvm.org/docs/LibFuzzer.html}}.


\bibitem[Liu et~al\mbox{.}(2023b)]%
        {nnsmith}
\bibfield{author}{\bibinfo{person}{Jiawei Liu}, \bibinfo{person}{Jinkun Lin}, \bibinfo{person}{Fabian Ruffy}, \bibinfo{person}{Cheng Tan}, \bibinfo{person}{Jinyang Li}, \bibinfo{person}{Aurojit Panda}, {and} \bibinfo{person}{Lingming Zhang}.} \bibinfo{year}{2023}\natexlab{b}.
\newblock \showarticletitle{NNSmith: Generating Diverse and Valid Test Cases for Deep Learning Compilers}. In \bibinfo{booktitle}{\emph{ASPLOS}}. \bibinfo{pages}{530–543}.
\newblock


\bibitem[Liu et~al\mbox{.}(2023c)]%
        {neuri}
\bibfield{author}{\bibinfo{person}{Jiawei Liu}, \bibinfo{person}{Jinjun Peng}, \bibinfo{person}{Yuyao Wang}, {and} \bibinfo{person}{Lingming Zhang}.} \bibinfo{year}{2023}\natexlab{c}.
\newblock \showarticletitle{NeuRI: Diversifying DNN Generation via Inductive Rule Inference}. In \bibinfo{booktitle}{\emph{Proceedings of the 31st ACM Joint European Software Engineering Conference and Symposium on the Foundations of Software Engineering}} (San Francisco, CA, USA) \emph{(\bibinfo{series}{ESEC/FSE 2023})}. \bibinfo{publisher}{Association for Computing Machinery}, \bibinfo{address}{New York, NY, USA}, \bibinfo{pages}{657--669}.
\newblock
\showISBNx{9798400703270}
\urldef\tempurl%
\url{https://doi.org/10.1145/3611643.3616337}
\showDOI{\tempurl}


\bibitem[Liu et~al\mbox{.}(2022)]%
        {tzer}
\bibfield{author}{\bibinfo{person}{Jiawei Liu}, \bibinfo{person}{Yuxiang Wei}, \bibinfo{person}{Sen Yang}, \bibinfo{person}{Yinlin Deng}, {and} \bibinfo{person}{Lingming Zhang}.} \bibinfo{year}{2022}\natexlab{}.
\newblock \showarticletitle{Coverage-Guided Tensor Compiler Fuzzing with Joint IR-Pass Mutation}.
\newblock \bibinfo{journal}{\emph{Proc. ACM Program. Lang.}} \bibinfo{volume}{6}, \bibinfo{number}{OOPSLA1}, Article \bibinfo{articleno}{73} (\bibinfo{date}{apr} \bibinfo{year}{2022}), \bibinfo{numpages}{26}~pages.
\newblock
\urldef\tempurl%
\url{https://doi.org/10.1145/3527317}
\showDOI{\tempurl}


\bibitem[Liu et~al\mbox{.}(2023a)]%
        {liu2023lost}
\bibfield{author}{\bibinfo{person}{Nelson~F Liu}, \bibinfo{person}{Kevin Lin}, \bibinfo{person}{John Hewitt}, \bibinfo{person}{Ashwin Paranjape}, \bibinfo{person}{Michele Bevilacqua}, \bibinfo{person}{Fabio Petroni}, {and} \bibinfo{person}{Percy Liang}.} \bibinfo{year}{2023}\natexlab{a}.
\newblock \showarticletitle{Lost in the middle: How language models use long contexts}.
\newblock \bibinfo{journal}{\emph{arXiv preprint arXiv:2307.03172}} (\bibinfo{year}{2023}).
\newblock


\bibitem[Liu and Meng(2023)]%
        {liu2023dsfuzz}
\bibfield{author}{\bibinfo{person}{Yinxi Liu} {and} \bibinfo{person}{Wei Meng}.} \bibinfo{year}{2023}\natexlab{}.
\newblock \showarticletitle{DSFuzz: Detecting Deep State Bugs with Dependent State Exploration}. In \bibinfo{booktitle}{\emph{Proceedings of the 2023 ACM SIGSAC Conference on Computer and Communications Security}}. \bibinfo{pages}{1242--1256}.
\newblock


\bibitem[Livinskii et~al\mbox{.}(2020)]%
        {yarpgen}
\bibfield{author}{\bibinfo{person}{Vsevolod Livinskii}, \bibinfo{person}{Dmitry Babokin}, {and} \bibinfo{person}{John Regehr}.} \bibinfo{year}{2020}\natexlab{}.
\newblock \showarticletitle{Random testing for C and C++ compilers with YARPGen}.
\newblock \bibinfo{journal}{\emph{Proceedings of the ACM on Programming Languages}} \bibinfo{volume}{4}, \bibinfo{number}{OOPSLA} (\bibinfo{year}{2020}), \bibinfo{pages}{1--25}.
\newblock


\bibitem[LLVM(2023)]%
        {llvmpass}
LLVM \bibinfo{year}{2023}\natexlab{}.
\newblock \bibinfo{title}{LLVM’s Analysis and Transform Passes}.
\newblock
\newblock
\newblock
\shownote{\url{https://llvm.org/docs/Passes.html}}.


\bibitem[Mathis et~al\mbox{.}(2020)]%
        {mathis2020learning}
\bibfield{author}{\bibinfo{person}{Bj{\"o}rn Mathis}, \bibinfo{person}{Rahul Gopinath}, {and} \bibinfo{person}{Andreas Zeller}.} \bibinfo{year}{2020}\natexlab{}.
\newblock \showarticletitle{Learning input tokens for effective fuzzing}. In \bibinfo{booktitle}{\emph{Proceedings of the 29th ACM SIGSOFT international symposium on software testing and analysis}}. \bibinfo{pages}{27--37}.
\newblock


\bibitem[McDonald and Xu(1995)]%
        {mcdonald1995generalization}
\bibfield{author}{\bibinfo{person}{James~B McDonald} {and} \bibinfo{person}{Yexiao~J Xu}.} \bibinfo{year}{1995}\natexlab{}.
\newblock \showarticletitle{A generalization of the beta distribution with applications}.
\newblock \bibinfo{journal}{\emph{Journal of Econometrics}} \bibinfo{volume}{66}, \bibinfo{number}{1-2} (\bibinfo{year}{1995}), \bibinfo{pages}{133--152}.
\newblock


\bibitem[McKeeman(1998)]%
        {mckeeman1998differential}
\bibfield{author}{\bibinfo{person}{William~M McKeeman}.} \bibinfo{year}{1998}\natexlab{}.
\newblock \showarticletitle{Differential testing for software}.
\newblock \bibinfo{journal}{\emph{Digital Technical Journal}} \bibinfo{volume}{10}, \bibinfo{number}{1} (\bibinfo{year}{1998}), \bibinfo{pages}{100--107}.
\newblock


\bibitem[Meng et~al\mbox{.}(2024)]%
        {meng2024large}
\bibfield{author}{\bibinfo{person}{Ruijie Meng}, \bibinfo{person}{Martin Mirchev}, \bibinfo{person}{Marcel B{\"o}hme}, {and} \bibinfo{person}{Abhik Roychoudhury}.} \bibinfo{year}{2024}\natexlab{}.
\newblock \showarticletitle{Large language model guided protocol fuzzing}. In \bibinfo{booktitle}{\emph{Proceedings of the 31st Annual Network and Distributed System Security Symposium (NDSS)}}.
\newblock


\bibitem[MKLDNN(2024)]%
        {mkldnn}
MKLDNN \bibinfo{year}{2024}\natexlab{}.
\newblock \bibinfo{title}{MKL-DNN}.
\newblock
\newblock
\newblock
\shownote{\url{https://github.com/rsdubtso/mkl-dnn}}.


\bibitem[Nie et~al\mbox{.}(2023)]%
        {teco}
\bibfield{author}{\bibinfo{person}{Pengyu Nie}, \bibinfo{person}{Rahul Banerjee}, \bibinfo{person}{Junyi~Jessy Li}, \bibinfo{person}{Raymond~J Mooney}, {and} \bibinfo{person}{Milos Gligoric}.} \bibinfo{year}{2023}\natexlab{}.
\newblock \showarticletitle{Learning Deep Semantics for Test Completion}.
\newblock \bibinfo{journal}{\emph{arXiv preprint arXiv:2302.10166}} (\bibinfo{year}{2023}).
\newblock


\bibitem[oneDNN(2024)]%
        {onednn}
oneDNN \bibinfo{year}{2024}\natexlab{}.
\newblock \bibinfo{title}{oneDNN}.
\newblock
\newblock
\newblock
\shownote{\url{https://github.com/oneapi-src/oneDNN}}.


\bibitem[OpenAI(2023a)]%
        {chatgpt}
\bibfield{author}{\bibinfo{person}{OpenAI}.} \bibinfo{year}{2023}\natexlab{a}.
\newblock \showarticletitle{ChatGPT}.
\newblock  (\bibinfo{year}{2023}).
\newblock
\newblock
\shownote{\url{https://openai.com/blog/chatgpt}}.


\bibitem[OpenAI(2023b)]%
        {openai2023gpt4}
\bibfield{author}{\bibinfo{person}{OpenAI}.} \bibinfo{year}{2023}\natexlab{b}.
\newblock \bibinfo{title}{GPT-4 Technical Report}.
\newblock
\newblock
\showeprint[arxiv]{2303.08774}~[cs.CL]


\bibitem[Ou et~al\mbox{.}(2024)]%
        {ou2024mutators}
\bibfield{author}{\bibinfo{person}{Xianfei Ou}, \bibinfo{person}{Cong Li}, \bibinfo{person}{Yanyan Jiang}, {and} \bibinfo{person}{Chang Xu}.} \bibinfo{year}{2024}\natexlab{}.
\newblock \showarticletitle{The Mutators Reloaded: Fuzzing Compilers with Large Language Model Generated Mutation Operators}.
\newblock  (\bibinfo{year}{2024}).
\newblock


\bibitem[Pei et~al\mbox{.}(2017)]%
        {pei2017deepxplore}
\bibfield{author}{\bibinfo{person}{Kexin Pei}, \bibinfo{person}{Yinzhi Cao}, \bibinfo{person}{Junfeng Yang}, {and} \bibinfo{person}{Suman Jana}.} \bibinfo{year}{2017}\natexlab{}.
\newblock \showarticletitle{Deepxplore: Automated whitebox testing of deep learning systems}. In \bibinfo{booktitle}{\emph{proceedings of the 26th Symposium on Operating Systems Principles}}. \bibinfo{pages}{1--18}.
\newblock


\bibitem[Pham et~al\mbox{.}(2019)]%
        {cradle}
\bibfield{author}{\bibinfo{person}{Hung~Viet Pham}, \bibinfo{person}{Thibaud Lutellier}, \bibinfo{person}{Weizhen Qi}, {and} \bibinfo{person}{Lin Tan}.} \bibinfo{year}{2019}\natexlab{}.
\newblock \showarticletitle{{CRADLE: Cross-Backend Validation to Detect and Localize Bugs in Deep Learning Libraries}}. In \bibinfo{booktitle}{\emph{2019 IEEE/ACM 41st International Conference on Software Engineering (ICSE)}}. \bibinfo{pages}{1027--1038}.
\newblock
\urldef\tempurl%
\url{https://doi.org/10.1109/ICSE.2019.00107}
\showDOI{\tempurl}


\bibitem[PyTorch(2023a)]%
        {pytorch}
PyTorch \bibinfo{year}{2023}\natexlab{a}.
\newblock \bibinfo{title}{PyTorch}.
\newblock
\newblock
\newblock
\shownote{\url{http://pytorch.org}}.


\bibitem[PyTorch(2023b)]%
        {pytorch2}
PyTorch \bibinfo{year}{2023}\natexlab{b}.
\newblock \bibinfo{title}{PyTorch 2.0}.
\newblock
\newblock
\newblock
\shownote{\url{https://pytorch.org/get-started/pytorch-2.0}}.


\bibitem[Sch{\"a}fer et~al\mbox{.}(2023)]%
        {testpilot}
\bibfield{author}{\bibinfo{person}{Max Sch{\"a}fer}, \bibinfo{person}{Sarah Nadi}, \bibinfo{person}{Aryaz Eghbali}, {and} \bibinfo{person}{Frank Tip}.} \bibinfo{year}{2023}\natexlab{}.
\newblock \showarticletitle{Adaptive test generation using a large language model}.
\newblock \bibinfo{journal}{\emph{arXiv preprint arXiv:2302.06527}} (\bibinfo{year}{2023}).
\newblock


\bibitem[Security(2007)]%
        {jsfunfuzz}
\bibfield{author}{\bibinfo{person}{Mozilla Security}.} \bibinfo{year}{2007}\natexlab{}.
\newblock \bibinfo{title}{jsfunfuzz}.
\newblock \bibinfo{howpublished}{\url{https://github.com/MozillaSecurity/funfuzz}}.
\newblock


\bibitem[Sen(2007)]%
        {sen2007concolic}
\bibfield{author}{\bibinfo{person}{Koushik Sen}.} \bibinfo{year}{2007}\natexlab{}.
\newblock \showarticletitle{Concolic testing}. In \bibinfo{booktitle}{\emph{Proceedings of the 22nd IEEE/ACM international conference on Automated software engineering}}. \bibinfo{pages}{571--572}.
\newblock


\bibitem[Sen et~al\mbox{.}(2005)]%
        {cute}
\bibfield{author}{\bibinfo{person}{Koushik Sen}, \bibinfo{person}{Darko Marinov}, {and} \bibinfo{person}{Gul Agha}.} \bibinfo{year}{2005}\natexlab{}.
\newblock \showarticletitle{CUTE: A Concolic Unit Testing Engine for C}. In \bibinfo{booktitle}{\emph{Proceedings of the 10th European Software Engineering Conference Held Jointly with 13th ACM SIGSOFT International Symposium on Foundations of Software Engineering}} (Lisbon, Portugal) \emph{(\bibinfo{series}{ESEC/FSE-13})}. \bibinfo{publisher}{Association for Computing Machinery}, \bibinfo{address}{New York, NY, USA}, \bibinfo{pages}{263–272}.
\newblock
\showISBNx{1595930140}
\urldef\tempurl%
\url{https://doi.org/10.1145/1081706.1081750}
\showDOI{\tempurl}


\bibitem[Shao et~al\mbox{.}(2023)]%
        {Shao2023AnES}
\bibfield{author}{\bibinfo{person}{Weijie Shao}, \bibinfo{person}{Yuyang Gao}, \bibinfo{person}{Fu Song}, \bibinfo{person}{Sen Chen}, {and} \bibinfo{person}{Lingling Fan}.} \bibinfo{year}{2023}\natexlab{}.
\newblock \showarticletitle{An Empirical Study of Bugs in Open-Source Federated Learning Framework}.
\newblock \bibinfo{journal}{\emph{ArXiv}}  \bibinfo{volume}{abs/2308.05014} (\bibinfo{year}{2023}).
\newblock
\urldef\tempurl%
\url{https://api.semanticscholar.org/CorpusID:265221980}
\showURL{%
\tempurl}


\bibitem[Shreiner et~al\mbox{.}(2009)]%
        {opengl}
\bibfield{author}{\bibinfo{person}{Dave Shreiner} {et~al\mbox{.}}} \bibinfo{year}{2009}\natexlab{}.
\newblock \bibinfo{booktitle}{\emph{OpenGL programming guide: the official guide to learning OpenGL, versions 3.0 and 3.1}}.
\newblock \bibinfo{publisher}{Pearson Education}.
\newblock


\bibitem[Su et~al\mbox{.}(2021)]%
        {su2021benchmarking}
\bibfield{author}{\bibinfo{person}{Ting Su}, \bibinfo{person}{Jue Wang}, {and} \bibinfo{person}{Zhendong Su}.} \bibinfo{year}{2021}\natexlab{}.
\newblock \showarticletitle{Benchmarking automated GUI testing for Android against real-world bugs}. In \bibinfo{booktitle}{\emph{Proceedings of the 29th ACM Joint Meeting on European Software Engineering Conference and Symposium on the Foundations of Software Engineering}}. \bibinfo{pages}{119--130}.
\newblock


\bibitem[Sun et~al\mbox{.}(2016)]%
        {sun2016toward}
\bibfield{author}{\bibinfo{person}{Chengnian Sun}, \bibinfo{person}{Vu Le}, \bibinfo{person}{Qirun Zhang}, {and} \bibinfo{person}{Zhendong Su}.} \bibinfo{year}{2016}\natexlab{}.
\newblock \showarticletitle{Toward understanding compiler bugs in GCC and LLVM}. In \bibinfo{booktitle}{\emph{Proceedings of the 25th international symposium on software testing and analysis}}. \bibinfo{pages}{294--305}.
\newblock


\bibitem[Sun et~al\mbox{.}(2023)]%
        {sunsmt}
\bibfield{author}{\bibinfo{person}{Maolin Sun}, \bibinfo{person}{Yibiao Yang}, \bibinfo{person}{Yang Wang}, \bibinfo{person}{Ming Wen}, \bibinfo{person}{Haoxiang Jia}, {and} \bibinfo{person}{Yuming Zhou}.} \bibinfo{year}{2023}\natexlab{}.
\newblock \showarticletitle{SMT Solver Validation Empowered by Large Pre-trained Language Models}. In \bibinfo{booktitle}{\emph{ASE}}.
\newblock


\bibitem[Sutton et~al\mbox{.}(2007)]%
        {sutton2007fuzzing}
\bibfield{author}{\bibinfo{person}{Michael Sutton}, \bibinfo{person}{Adam Greene}, {and} \bibinfo{person}{Pedram Amini}.} \bibinfo{year}{2007}\natexlab{}.
\newblock \bibinfo{booktitle}{\emph{Fuzzing: brute force vulnerability discovery}}.
\newblock \bibinfo{publisher}{Pearson Education}.
\newblock


\bibitem[Tang et~al\mbox{.}(2024)]%
        {tang2024chatgpt}
\bibfield{author}{\bibinfo{person}{Yutian Tang}, \bibinfo{person}{Zhijie Liu}, \bibinfo{person}{Zhichao Zhou}, {and} \bibinfo{person}{Xiapu Luo}.} \bibinfo{year}{2024}\natexlab{}.
\newblock \showarticletitle{Chatgpt vs sbst: A comparative assessment of unit test suite generation}.
\newblock \bibinfo{journal}{\emph{IEEE Transactions on Software Engineering}} (\bibinfo{year}{2024}).
\newblock


\bibitem[TensorFlow(2023)]%
        {Tensorflow}
TensorFlow \bibinfo{year}{2023}\natexlab{}.
\newblock \bibinfo{title}{TensorFlow}.
\newblock
\newblock
\newblock
\shownote{\url{https://www.tensorflow.org}}.


\bibitem[TensorFlowLite(2023)]%
        {tflite}
TensorFlowLite \bibinfo{year}{2023}\natexlab{}.
\newblock \bibinfo{title}{TensorFlow Lite}.
\newblock
\newblock
\newblock
\shownote{\url{https://www.tensorflow.org/lite}}.


\bibitem[TensorFlowXLA(2023)]%
        {tfxla}
TensorFlowXLA \bibinfo{year}{2023}\natexlab{}.
\newblock \bibinfo{title}{TensorFlow XLA}.
\newblock
\newblock
\newblock
\shownote{\url{https://www.tensorflow.org/xla}}.


\bibitem[Thompson(1933)]%
        {thompson1933likelihood}
\bibfield{author}{\bibinfo{person}{William~R Thompson}.} \bibinfo{year}{1933}\natexlab{}.
\newblock \showarticletitle{On the likelihood that one unknown probability exceeds another in view of the evidence of two samples}.
\newblock \bibinfo{journal}{\emph{Biometrika}} \bibinfo{volume}{25}, \bibinfo{number}{3-4} (\bibinfo{year}{1933}), \bibinfo{pages}{285--294}.
\newblock


\bibitem[Triton(2024)]%
        {triton}
Triton \bibinfo{year}{2024}\natexlab{}.
\newblock \bibinfo{title}{Triton}.
\newblock
\newblock
\newblock
\shownote{\url{https://github.com/openai/triton}}.


\bibitem[Vaswani et~al\mbox{.}(2017)]%
        {vaswani2017attention}
\bibfield{author}{\bibinfo{person}{Ashish Vaswani}, \bibinfo{person}{Noam Shazeer}, \bibinfo{person}{Niki Parmar}, \bibinfo{person}{Jakob Uszkoreit}, \bibinfo{person}{Llion Jones}, \bibinfo{person}{Aidan~N Gomez}, \bibinfo{person}{{\L}ukasz Kaiser}, {and} \bibinfo{person}{Illia Polosukhin}.} \bibinfo{year}{2017}\natexlab{}.
\newblock \showarticletitle{Attention is all you need}.
\newblock \bibinfo{journal}{\emph{Advances in neural information processing systems}}  \bibinfo{volume}{30} (\bibinfo{year}{2017}).
\newblock


\bibitem[Wang et~al\mbox{.}(2022)]%
        {eagle}
\bibfield{author}{\bibinfo{person}{Jiannan Wang}, \bibinfo{person}{Thibaud Lutellier}, \bibinfo{person}{Shangshu Qian}, \bibinfo{person}{Hung~Viet Pham}, {and} \bibinfo{person}{Lin Tan}.} \bibinfo{year}{2022}\natexlab{}.
\newblock \showarticletitle{EAGLE: Creating Equivalent Graphs to Test Deep Learning Libraries}.
\newblock  (\bibinfo{year}{2022}).
\newblock


\bibitem[Wang et~al\mbox{.}(2020)]%
        {lemon}
\bibfield{author}{\bibinfo{person}{Zan Wang}, \bibinfo{person}{Ming Yan}, \bibinfo{person}{Junjie Chen}, \bibinfo{person}{Shuang Liu}, {and} \bibinfo{person}{Dongdi Zhang}.} \bibinfo{year}{2020}\natexlab{}.
\newblock \showarticletitle{Deep learning library testing via effective model generation}. In \bibinfo{booktitle}{\emph{Proceedings of the 28th ACM Joint Meeting on European Software Engineering Conference and Symposium on the Foundations of Software Engineering}}. \bibinfo{pages}{788--799}.
\newblock


\bibitem[Wei et~al\mbox{.}(2022)]%
        {freefuzz}
\bibfield{author}{\bibinfo{person}{Anjiang Wei}, \bibinfo{person}{Yinlin Deng}, \bibinfo{person}{Chenyuan Yang}, {and} \bibinfo{person}{Lingming Zhang}.} \bibinfo{year}{2022}\natexlab{}.
\newblock \showarticletitle{Free Lunch for Testing: Fuzzing Deep-Learning Libraries from Open Source}. In \bibinfo{booktitle}{\emph{2022 IEEE/ACM 44th International Conference on Software Engineering (ICSE)}}. \bibinfo{pages}{995--1007}.
\newblock
\urldef\tempurl%
\url{https://doi.org/10.1145/3510003.3510041}
\showDOI{\tempurl}


\bibitem[Wei et~al\mbox{.}(2024)]%
        {wei2024magicoder}
\bibfield{author}{\bibinfo{person}{Yuxiang Wei}, \bibinfo{person}{Zhe Wang}, \bibinfo{person}{Jiawei Liu}, \bibinfo{person}{Yifeng Ding}, {and} \bibinfo{person}{Lingming Zhang}.} \bibinfo{year}{2024}\natexlab{}.
\newblock \showarticletitle{Magicoder: Empowering code generation with oss-instruct}. In \bibinfo{booktitle}{\emph{Forty-first International Conference on Machine Learning}}.
\newblock


\bibitem[Xia et~al\mbox{.}(2023)]%
        {xia2023universal}
\bibfield{author}{\bibinfo{person}{Chunqiu~Steven Xia}, \bibinfo{person}{Matteo Paltenghi}, \bibinfo{person}{Jia~Le Tian}, \bibinfo{person}{Michael Pradel}, {and} \bibinfo{person}{Lingming Zhang}.} \bibinfo{year}{2023}\natexlab{}.
\newblock \showarticletitle{Universal fuzzing via large language models}.
\newblock \bibinfo{journal}{\emph{arXiv preprint arXiv:2308.04748}} (\bibinfo{year}{2023}).
\newblock


\bibitem[Xie et~al\mbox{.}(2022)]%
        {docter}
\bibfield{author}{\bibinfo{person}{Danning Xie}, \bibinfo{person}{Yitong Li}, \bibinfo{person}{Mijung Kim}, \bibinfo{person}{Hung~Viet Pham}, \bibinfo{person}{Lin Tan}, \bibinfo{person}{Xiangyu Zhang}, {and} \bibinfo{person}{Michael~W Godfrey}.} \bibinfo{year}{2022}\natexlab{}.
\newblock \showarticletitle{DocTer: documentation-guided fuzzing for testing deep learning API functions}. In \bibinfo{booktitle}{\emph{Proceedings of the 31st ACM SIGSOFT International Symposium on Software Testing and Analysis}}. \bibinfo{pages}{176--188}.
\newblock


\bibitem[Xu et~al\mbox{.}(2023)]%
        {xu2023silent}
\bibfield{author}{\bibinfo{person}{Jianhao Xu}, \bibinfo{person}{Kangjie Lu}, \bibinfo{person}{Zhengjie Du}, \bibinfo{person}{Zhu Ding}, \bibinfo{person}{Linke Li}, \bibinfo{person}{Qiushi Wu}, \bibinfo{person}{Mathias Payer}, {and} \bibinfo{person}{Bing Mao}.} \bibinfo{year}{2023}\natexlab{}.
\newblock \showarticletitle{Silent Bugs Matter: A Study of $\{$Compiler-Introduced$\}$ Security Bugs}. In \bibinfo{booktitle}{\emph{32nd USENIX Security Symposium (USENIX Security 23)}}. \bibinfo{pages}{3655--3672}.
\newblock


\bibitem[Yang et~al\mbox{.}(2023a)]%
        {nablafuzz}
\bibfield{author}{\bibinfo{person}{Chenyuan Yang}, \bibinfo{person}{Yinlin Deng}, \bibinfo{person}{Jiayi Yao}, \bibinfo{person}{Yuxing Tu}, \bibinfo{person}{Hanchi Li}, {and} \bibinfo{person}{Lingming Zhang}.} \bibinfo{year}{2023}\natexlab{a}.
\newblock \showarticletitle{Fuzzing Automatic Differentiation in Deep-Learning Libraries}. In \bibinfo{booktitle}{\emph{Proceedings of the 45th International Conference on Software Engineering}} (Melbourne, Victoria, Australia) \emph{(\bibinfo{series}{ICSE '23})}. \bibinfo{publisher}{IEEE Press}, \bibinfo{pages}{1174–1186}.
\newblock
\showISBNx{9781665457019}
\urldef\tempurl%
\url{https://doi.org/10.1109/ICSE48619.2023.00105}
\showDOI{\tempurl}


\bibitem[Yang et~al\mbox{.}(2023b)]%
        {yang2023kernelgpt}
\bibfield{author}{\bibinfo{person}{Chenyuan Yang}, \bibinfo{person}{Zijie Zhao}, {and} \bibinfo{person}{Lingming Zhang}.} \bibinfo{year}{2023}\natexlab{b}.
\newblock \showarticletitle{Kernelgpt: Enhanced kernel fuzzing via large language models}.
\newblock \bibinfo{journal}{\emph{arXiv preprint arXiv:2401.00563}} (\bibinfo{year}{2023}).
\newblock


\bibitem[Yang et~al\mbox{.}(2011)]%
        {csmith}
\bibfield{author}{\bibinfo{person}{Xuejun Yang}, \bibinfo{person}{Yang Chen}, \bibinfo{person}{Eric Eide}, {and} \bibinfo{person}{John Regehr}.} \bibinfo{year}{2011}\natexlab{}.
\newblock \showarticletitle{Finding and understanding bugs in C compilers}. In \bibinfo{booktitle}{\emph{Proceedings of the 32nd ACM SIGPLAN conference on Programming language design and implementation}}. \bibinfo{pages}{283--294}.
\newblock


\bibitem[Yang et~al\mbox{.}(2017)]%
        {yang2017dead}
\bibfield{author}{\bibinfo{person}{Zhaomo Yang}, \bibinfo{person}{Brian Johannesmeyer}, \bibinfo{person}{Anders~Trier Olesen}, \bibinfo{person}{Sorin Lerner}, {and} \bibinfo{person}{Kirill Levchenko}.} \bibinfo{year}{2017}\natexlab{}.
\newblock \showarticletitle{Dead store elimination (still) considered harmful}. In \bibinfo{booktitle}{\emph{26th USENIX Security Symposium (USENIX Security 17)}}. \bibinfo{pages}{1025--1040}.
\newblock


\bibitem[Yue~Wang and Hoi(2021)]%
        {wang2021codet5}
\bibfield{author}{\bibinfo{person}{Shafiq~Joty Yue~Wang, Weishi~Wang} {and} \bibinfo{person}{Steven~C.H. Hoi}.} \bibinfo{year}{2021}\natexlab{}.
\newblock \showarticletitle{CodeT5: Identifier-aware Unified Pre-trained Encoder-Decoder Models for Code Understanding and Generation}. In \bibinfo{booktitle}{\emph{EMNLP 2021}}.
\newblock


\bibitem[Yun et~al\mbox{.}(2018)]%
        {qsym}
\bibfield{author}{\bibinfo{person}{Insu Yun}, \bibinfo{person}{Sangho Lee}, \bibinfo{person}{Meng Xu}, \bibinfo{person}{Yeongjin Jang}, {and} \bibinfo{person}{Taesoo Kim}.} \bibinfo{year}{2018}\natexlab{}.
\newblock \showarticletitle{$\{$QSYM$\}$: A practical concolic execution engine tailored for hybrid fuzzing}. In \bibinfo{booktitle}{\emph{27th USENIX Security Symposium (USENIX Security 18)}}. \bibinfo{pages}{745--761}.
\newblock


\bibitem[Zeller et~al\mbox{.}(2019)]%
        {zeller2019fuzzing}
\bibfield{author}{\bibinfo{person}{Andreas Zeller}, \bibinfo{person}{Rahul Gopinath}, \bibinfo{person}{Marcel B{\"o}hme}, \bibinfo{person}{Gordon Fraser}, {and} \bibinfo{person}{Christian Holler}.} \bibinfo{year}{2019}\natexlab{}.
\newblock \bibinfo{title}{The fuzzing book}.
\newblock
\newblock


\bibitem[Zhang et~al\mbox{.}(2023)]%
        {zhang2023survey}
\bibfield{author}{\bibinfo{person}{Quanjun Zhang}, \bibinfo{person}{Chunrong Fang}, \bibinfo{person}{Yang Xie}, \bibinfo{person}{Yaxin Zhang}, \bibinfo{person}{Yun Yang}, \bibinfo{person}{Weisong Sun}, \bibinfo{person}{Shengcheng Yu}, {and} \bibinfo{person}{Zhenyu Chen}.} \bibinfo{year}{2023}\natexlab{}.
\newblock \showarticletitle{A survey on large language models for software engineering}.
\newblock \bibinfo{journal}{\emph{arXiv preprint arXiv:2312.15223}} (\bibinfo{year}{2023}).
\newblock


\bibitem[Zhang et~al\mbox{.}(2017)]%
        {zhang2017skeletal}
\bibfield{author}{\bibinfo{person}{Qirun Zhang}, \bibinfo{person}{Chengnian Sun}, {and} \bibinfo{person}{Zhendong Su}.} \bibinfo{year}{2017}\natexlab{}.
\newblock \showarticletitle{Skeletal program enumeration for rigorous compiler testing}. In \bibinfo{booktitle}{\emph{Proceedings of the 38th ACM SIGPLAN Conference on Programming Language Design and Implementation}}. \bibinfo{pages}{347--361}.
\newblock


\end{thebibliography}
